\documentclass[10pt]{article}

\usepackage{amsmath}
\usepackage{amssymb}
\usepackage{latexsym}
\usepackage{graphicx}
\usepackage{epsfig}
\usepackage{stmaryrd}
\usepackage{framed}

\newtheorem{thm}{\bf Theorem}[section]
\newtheorem{lem}[thm]{\bf Lemma}        
\newtheorem{prop}[thm]{\bf Proposition}  
\newtheorem{cor}[thm]{\bf Corollary}

\newtheorem{expl}[thm]{\bf Example}       
\newtheorem{remark}[thm]{\bf Remark}

\newcommand{\Mod}{\mathbf{Mod}}
\newcommand{\SMod}{\mathbf{SMod}}
\newcommand{\SymMod}{\mathbf{SymMod}}
\newcommand{\SymSMod}{\mathbf{SymSMod}}
\newcommand{\id}{\mathbf{id}}
\newcommand{\one}{\mathbf{1}}

\def\Bbox{
{\unskip\nobreak\hfil\penalty50
\hskip1em\hbox{}\nobreak\hfil{\lower .5pt \hbox{$\Box$}}
\parfillskip=0pt \finalhyphendemerits=0 \par}
}

\def\eop{
\ifmmode {\hbox{\Bbox}} \else \Bbox \fi
}

\def\bbox{
\ifmmode {\hbox{\bbox}} \else \Bbox \fi
}

\begin{document}

\title{\bf Equational properties of stratified least fixed points}

\author{Zolt\'an \'Esik\thanks{Partially supported by 
grant no. ANN 110883 from the National Foundation of Hungary for 
Scientific Research.}\\
Dept. of Computer Science\\
University of Szeged\\
Hungary}

\date{\empty}

\maketitle

\begin{abstract}
Recently, a novel fixed point operation has been introduced 
over certain non-monotonic functions between stratified complete lattices
and used to give semantics to logic programs with negation 
and boolean context-free grammars. We prove that this
new operation satisfies `the standard' identities of 
fixed point operations as described by the axioms of 
iteration theories.
We also study this new fixed point operation in connection
with lambda-abstraction.  
\end{abstract}

\section{Introduction}

The semantics of negation free logic programs is classically defined as 
the least fixed point of the `immediate consequence operation' canonically associated with the program,
cf. \cite{EmdenKowalski}. Since this operation is monotonic, the existence of the 
least fixed point is guaranteed by the well-known Knaster-Tarski 
theorem \cite{Tarski}. However, for programs with negation,
the immediate consequence operation is not necessarily monotonic
and fixed points are not guaranteed to exist. The well-founded semantics
\cite{Gelder,Prz} of logic programs with negation is based on a three-valued
(or sometimes four-valued) logic and defines the semantics of a program 
as the least fixed point of the so-called `stable operation' associated with 
the program with respect to the information, or knowledge, or Fitting ordering \cite{Fitting}. 
The well-founded approach has led to the development of a deep abstract 
fixed point theory for non-monotonic functions which in turn has successfully been applied
to problems in various areas beyond logic programming, 
see \cite{Deneckeretal1,Deneckeretal,Fitting,Vennekensetal}
for a sampling of articles covering such results. 

Another approach to the semantics of logic programs with negation
based on an infinite structure of truth values was introduced in \cite{RW}. 
It has been demonstrated that the immediate consequence operation 
associated with a logic program has a unique minimum model
with respect to a novel ordering of the possible
interpretations of the program variables over 
the truth values. An advantage of this approach is that
it uses the immediate consequence operation in a direct way.  
A disadvantage is that it relies on a more complex logic of truth values. However, it 
does provide more information about the level of certainty 
of truth or falsity. The development of an abstract fixed point 
theory underlying the infinite valued approach has recently been undertaken 
in \cite{ERfp,ERwollic}. In \cite{ERfp}, certain stratified complete
lattices --called models-- were defined, consisting of a 
complete lattice $(L,\leq)$ and a family 
$(\sqsubseteq_\alpha)_{\alpha < \kappa}$ of preorderings indexed by
the ordinals $\alpha$ less than a fixed nonzero ordinal $\kappa$. 
Several axioms were imposed on models relating the lattice order $\leq$ to the 
preorderings $\sqsubseteq_\alpha$. It was established that in such models the 
preorderings $\sqsubseteq_\alpha$ determine another complete 
lattice structure $(L,\sqsubseteq)$, and that 
if an endofunction of a model satisfies some weak
monotonicity or continuity property (it is $\alpha$-monotonic or $\alpha$-continuous
for each ordinal $\alpha < \kappa$), then it has a least pre-fixed point with 
respect to the ordering $\sqsubseteq$, which is a fixed point. 
(These functions are not necessarily monotonic w.r.t. the ordering $\sqsubseteq$.)
This fixed point theorem has been applied to higher order logic 
programs and boolean grammars, cf. \cite{CERhigher,ERwollic}. 

A general study of the equational properties of fixed point operations 
in the context of Lawvere theories or the slightly more general 
cartesian categories has been provided in \cite{BEbook}. Several other formalisms 
may also be used for the same purpose including abstract clones or 
 $\mu$-expressions, or let-rec expressions. 
 It has been shown that the major fixed point operations  
commonly used in computer science, including the least fixed
point operation over monotonic or continuous functions between 
complete lattices or cpo's, or in continuous or 
rational theories \cite{ADJ},  
the unique fixed point operation over contractive functions of complete metric spaces 
or in Elgot's iterative theories \cite{Elgot},  
the initial fixed point operation over continuous functors 
over certain categories or in algebraically complete
categories, and many other models share the equational 
properties captured by the axioms of iteration theories, 
or iteration categories, cf.  
\cite{BEcircular,Esaxioms,EsikLabella} and \cite{BEbook}.
For a recent survey, see \cite{EsMFCS2015}.

Our main aim in this paper is to show that, in conjunction with 
the cartesian structure,  the new fixed point operation
introduced in \cite{ERfp} also satisfies the axioms of iteration 
theories. It then \emph{follows} that an identity is satisfied by
the new fixed point operation iff it holds in all 
iteration theories. 
We also define cartesian closed categories of models 
and establish the abstraction identity introduced in \cite{BEccc} 
that connects the fixed point operation to lambda abstraction. 

The paper is organized as follows. In Section~\ref{sec-models}, we recall
the axioms of models 
from \cite{ERfp}. We point out that models and $\alpha$-monotonic or
$\alpha$-continuous functions form cartesian categories
denoted $\Mod_m$ and $\Mod_c$.
Then, using the fixed point theorem of \cite{ERfp}, 
in Section~\ref{sec-fp} we endow $\Mod_m$ and 
$\Mod_c$ with a (parametrized) fixed point operation. 
Section~\ref{sec-cartesian} is devoted to proving that the 
identities of iteration theories hold in both categories
$\Mod_m$ and $\Mod_c$. In Section~\ref{sec-abstraction} 
we introduce cartesian closed categories of models and 
establish the abstraction identity 
that connects the fixed point operation to lambda abstraction. 
In Section~\ref{sec-variations} we consider several subcategories 
of $\Mod_m$ and $\Mod_c$. 

\emph{Some notation.} Composition of morphisms $f: L \to L'$ and $g: L' \to L''$
in any category is denoted $g \circ f$. The identity morphism associated with 
an object $L$ is denoted $\id_L$. Function application is written $f(x)$,
or just $fx$.

\section{Stratified complete lattices}
\label{sec-models}

Let $\kappa$ be a fixed nonzero ordinal, typically a limit ordinal. 
In \cite{ERfp}, stratified complete lattices equipped with a family 
of preorderings indexed by the ordinals $\alpha < \kappa$ subject to 
certain axioms were considered. 
In this section we recall the definition of these structures.
Following \cite{ERfp}, we also define $\alpha$-monotonic and 
$\alpha$-continuous functions between models 
and prove that they yield cartesian categories (cc's). 
For elementary facts about categories, the reader is referred to
\cite{BarrWells}. 

Suppose that $(L,\leq)$ is a complete lattice \cite{Daveyetal} with least and greatest 
elements $\bot$ and $\top$ which is equipped with a 
family $(\sqsubseteq_\alpha)_{\alpha < \kappa}$ of preorderings. For each 
$\alpha < \kappa$, let $ =_\alpha$ denote the equivalence relation determined 
by $\sqsubseteq_\alpha$, so that $x =_\alpha y$ 
iff $x \sqsubseteq_\alpha y$ and $y \sqsubseteq_\alpha x$,  
for all $x,y \in L$.  We say that $(L,\leq, (\sqsubseteq_\alpha)_{\alpha < \kappa})$ is 
a \emph{stratified complete lattice}, or a {\em model}, for short,  if the following two axioms hold.

\emph{Ax1. For all $\alpha < \beta <\kappa$, $\sqsubseteq_\beta$ is included in $=_\alpha$, 
i.e., for all $x,y \in L$, if $x \sqsubseteq_\beta y$ then $x =_\alpha y$.}

\emph{Ax2. The intersection $\bigcap_{\alpha <\kappa} =_\alpha$ of the relations $=_\alpha$ 
is the equality relation,
so that if $x ,y \in L$ with $x=_\alpha y$ for all $\alpha < \kappa$, then 
$x = y$.}

Thus, for all $x,y \in L$, we have $x = y$ iff $x =_\alpha y$ for all $\alpha < \kappa$. 
Moreover, we say that a stratified complete  lattice $(L,\leq, (\sqsubseteq_\alpha)_{\alpha < \kappa})$ 
is a \emph{model}, if it additionally satisfies the following two axioms, where 
$(x]_\alpha  = \{y : \forall \beta < \alpha\ x =_\beta y\}$ for all $x \in L$. 

\emph{Ax3. For every $\alpha < \kappa$, $x \in L$ and $X \subseteq (x]_\alpha$,  
there exists some $z\in (x]_\alpha$ with the following  properties: 
\begin{itemize}
\item $X \sqsubseteq_\alpha z$, i.e., $y \sqsubseteq_\alpha z$ for all $y\in X$,
\item for all $y\in (x]_\alpha$, if $X \sqsubseteq_\alpha y$ then $z \sqsubseteq_\alpha y$ and $z \leq y$.
\end{itemize}}

\emph{Ax4. For every $\alpha < \kappa$ and nonempty $X \subseteq L$ and $y \in L$, 
if $X =_\alpha y$ (ie., $x =_\alpha y$ for all $x \in X$), then $\bigvee X =_\alpha y$.}

\begin{expl} \cite{ERfp,RW}
\label{expl-VZ}
Consider the following linearly ordered set $V$ of truth values:
$$F_0 < F_1< \cdots < F_\alpha < \cdots < 0 < \cdots < T_\alpha < \cdots < T_1 < T_0,$$
where $\alpha < \Omega$, the first uncountable ordinal. 
Let $Z$ denote a nonempty set of (propositional) variables and consider the set $L = V^Z$ 
equipped with the pointwise ordering, so that for all $f,g \in L$,
$f \leq g$ iff $fz \leq gz$ for all $z \in Z$. Then $(L,\leq)$ is a complete lattice. 
For each $f,g \in L$ and $\alpha < \Omega$, define $f \sqsubseteq_\alpha g$ 
iff for all $z\in Z$,
\begin{itemize}
\item $\forall \beta < \alpha$ $(f z = F_\beta \Leftrightarrow  g z = F_\beta$\ $\wedge$\ 
   $fz = T_\beta \Leftrightarrow gz = T_\beta)$,
\item $g z = F_\alpha \Rightarrow f z = F_\alpha \wedge f z = T_\alpha \Rightarrow 
g z = T_\alpha$. 
\end{itemize}
Then $L$ is a model. The intuition for the definition of the relations $\sqsubseteq_\alpha$ 
is that $f \sqsubseteq_\alpha g$ iff $f$ and $g$ agree below `stratum' $\alpha$, 
and at stratum $\alpha$ and above, $f$ is either below or equivalent to $g$ in the sense that if for some $z$, 
$fz = T_\alpha$, then $gz = T_\alpha$ and if $gz = F_\alpha$ then $fz = F_\alpha$. 

It is clear that the first two axioms hold. To see that $Ax3$ holds, suppose 
that $\alpha < \Omega$, $g \in V^Z$ and $G \subseteq (g]_\alpha$. Then 
for all $f \in G$, $fz = gz$ whenever $gz$ is in 
$\{F_\beta, T_\beta: \beta < \alpha\}$.
The function $h = \bigsqcup_\alpha G$ is given by $hz = gz$ if 
$gz  \in \{F_\beta, T_\beta: \beta < \alpha\}$. If this condition does not hold, 
then $hz = T_\alpha$ if there exists $f \in G$ with $fz = T_\alpha$,
$hz = F_\alpha$ if $fz = F_\alpha$ for all $f \in G$, 
and $hz = F_{\alpha + 1}$ otherwise.

Finally, $Ax4$ holds since if $H \subseteq V^Z$ is a nonempty set 
and $g \in V^Z$ and $\alpha < \Omega$ such that $fz = gz$ for all $f \in H$ 
and $z \in Z$ when $fz$ or $gz$ is in $\{F_\beta,T_\beta : \beta \leq \alpha\}$, then also $(\bigvee H)z = gz$
whenever $gz$ or $(\bigvee H)z$ is in $\{F_\beta,T_\beta : \beta \leq \alpha\}$.
\end{expl}

\begin{expl}
\label{expl-complete lattices}
Every complete lattice $(L,\leq)$ gives rise to a model.
Indeed, define $\sqsubseteq_0$ to be the relation $\leq$,
and for each $0 < \alpha< \kappa$, define the relation 
$\sqsubseteq_\alpha$ as the equality relation $=$.
\end{expl} 

\begin{expl}
\label{expl-compl2}
Suppose that both $(L,\leq)$ and $(L,\sqsubseteq_0)$ 
are complete lattices. When $0 < \alpha < \kappa$, let 
$\sqsubseteq_\alpha$ be the equality relation $=$. 
As before, for each $X \subseteq L$, let $\bigvee X$ 
denote the supremum of $X$ w.r.t. $\leq$. Moreover, 
let $\bigsqcup_0 X$ denote the supremum of $X$ 
w.r.t. $\sqsubseteq_0$. Then $L$ is a model iff one of 
the following three conditions holds: 
\begin{enumerate}
\item For all $X \subseteq L$ and $y \in L$, if $X \sqsubseteq_0 y$ then 
     $\bigsqcup_0 X \leq y$. 
\item  For all $x,y \in L$, if $x \sqsubseteq_0 y$ then $x \leq y$. 
\item For all $X \subseteq L$, $\bigvee X \leq \bigsqcup_0 X$. 
\end{enumerate} 
Indeed, these conditions are equivalent. 
If the first condition holds, then for all $x,y$, 
if $x \sqsubseteq_0 y$ then $x = \bigsqcup_0 \{x\} \leq y$. 
Suppose that the second condition holds. Then for all $X$, 
every upper bound of $X$ w.r.t. $\sqsubseteq_0$ is an upper bound of $X$ w.r.t. $\leq$, 
hence $\bigvee X \leq \bigsqcup_0 X$. And if this holds and 
$x \sqsubseteq_0 y$, then $\bigvee \{x,y\} \leq \bigsqcup_0 \{x,y\} = y$, 
hence $x \leq y$. Thus, the second and third conditions are equivalent. 
Assume finally that the second and third conditions hold and $X \sqsubseteq_0 y$. 
Then $\bigsqcup_0 X \sqsubseteq_0 y$ and thus $\bigsqcup_0 X \leq y$.

Now, axioms $Ax1-Ax4$  clearly hold with the 
possible exception of $Ax3$ for $\alpha = 0$. 
But if the above three equivalent conditions are satisfied, 
then $Ax3$ holds for $\alpha = 0$, since 
for any $X$,  $\bigsqcup_0 X$ is just that element whose 
existence is required in $Ax3$. Finally, if $L$ is 
a model, then the first of the three conditions holds by $Ax3$. 
\end{expl} 

Below we will often denote a model 
$(L,\leq,(\sqsubseteq_\alpha)_{\alpha< \kappa})$ 
by just $L$.

\begin{remark}
{\rm 
It is clear that in any model $L$, the element $z$ in $Ax3$ is uniquely determined by $x$, $X$ and $\alpha$, and in fact by $X$ and $\alpha$ if $X$ is not empty. 
We denote it by $\bigsqcup_\alpha X$ and will freely use this notation 
without specifying $x$ whenever $X$ is not empty. 
Given $X$ and $\alpha$, there exists some $x$ with $X \subseteq (x]_\alpha$
iff $y =_\beta y'$ for all $y,y'\in X$ and $\beta < \alpha$, and
when $X$ is not empty, 
$\bigsqcup_\alpha X$ is the unique element $z$ with $X \sqsubseteq_\alpha z$ and such that 
for all $y$ with $X \sqsubseteq_\alpha y$, 
both $z \sqsubseteq_\alpha y$ and $z \leq y$. 
However, when $X$ is empty, then $\bigsqcup_\alpha X$ depends on $x$, 
at least when $\alpha > 0$, since it is required to be 
 in $ (x]_\alpha$. When $X = \emptyset$, 
$\bigsqcup_\alpha X$ is the unique element $z$ with $z =_\beta x$ 
for all $\beta < \alpha$ and such that whenever $y =_\beta x$ for all 
$\beta < \alpha$, then $z \sqsubseteq_\alpha y$ and $z \leq y$. 
Hence $\bigsqcup_\alpha \emptyset$ is both the $\leq$-least element of $(x]_\alpha$ 
and a $\sqsubseteq_\alpha$-least element of $(x]_\alpha$. 
}
\end{remark} 

Note that when $L$ is a model, $x \in L$ and $\alpha < \kappa$, then
$\bigsqcup_\alpha \{x\}$ is the $\leq$-least $y \in L$ with $x \sqsubseteq_\alpha y$
and also the $\leq$-least $y\in L$ with $x =_\alpha y$, i.e., 
$\bigsqcup_\alpha \{x\}$ is the $\leq$-least element of $[x]_\alpha = \{y: y =_\alpha x\}$.  
Below we will use the notation $x|_\alpha = \bigsqcup_\alpha \{x\}$ 
for all $x \in L$ and $\alpha < \kappa$. For example, 
$\bot|_\alpha = \bot$ for all $\alpha < \kappa$, since $\bot =_\alpha \bot$ 
and $\bot \leq x$ for all $x\in L$. 
By the above, $x =_\alpha x|_\alpha$ 
and $x =_\alpha y$ iff $x|_\alpha =_\alpha y|_\alpha$ iff $x|_\alpha = y|_\alpha$ 
for all $x,y \in X$ and $\alpha < \kappa$. 
It then follows by $Ax2$ that for all $x,y\in L$, $x = y$ iff 
$x|_\alpha = y|_\alpha$ for all $\alpha < \kappa$.
Moreover, if $x \in L$ and $\alpha + 1 < \kappa$
(which always holds when $\alpha < \kappa$ and $\kappa$ is a limit ordinal), 
then $x|_\alpha$ is the $\sqsubseteq_{\alpha+1}$-least element of $[x]_\alpha$. 
Indeed, $x|_\alpha$ is the $\leq$-least element of $[x]_\alpha$, whereas the $\leq$-least element of $(x]_{\alpha + 1}$ 
is $\bigsqcup_{\alpha + 1} \emptyset$ (with $\emptyset$ considered as a subset of $(x]_{\alpha + 1}$). 
However, $[x]_\alpha = (x]_{\alpha + 1}$, so the two least elements are equal. 
(See also Lemma 3.7 in \cite{ERfp}.)

It is known (see Lemma 3.12 in \cite{ERfp}) that the following conditions 
are equivalent for each $x \in L$ and $\alpha < \kappa$:
\begin{itemize}
\item $x = \bigsqcup_\alpha \{x\}$,
\item there exists $y \in L$ with $x = \bigsqcup_\alpha \{y\}$,
\item there exists a (nonempty) $X \subseteq L$ with $x = \bigsqcup_\alpha X$. 
\end{itemize}
 For later use, we prove:

\begin{lem}
\label{lem-min}
Suppose that $L$ is a model, $x \in L$ and $\alpha,\beta < \kappa$. 
Then $(x|_\alpha)|_\beta = x|_{\min\{\alpha,\beta\}}$. 
\end{lem}

{\sl Proof.} Suppose first that $\alpha \leq \beta$. We clearly have  
$x|_\alpha =_\beta (x|_\alpha)|_\beta$.
Moreover, if $x|_\alpha \sqsubseteq_\beta y$ then $x =_\alpha x|_\alpha \sqsubseteq_\alpha y $ by $Ax1$,
hence $x|_\alpha \leq y$. We conclude that $x|_\alpha = (x|_\alpha)|_\beta$. 
Now let $\beta < \alpha$. Since $x =_\alpha x|_\alpha$ and $\beta < \alpha$,
by $Ax1$ it holds that $x =_\beta x|_\alpha$. Hence for all $y \in L$, 
we have $x \sqsubseteq_\beta y$ iff $x|_\alpha \sqsubseteq_\beta y$. 
It follows that $x|_\beta = (x|_\alpha)|_\beta$. \eop

\begin{lem}
\label{lem-assoc}
Suppose that $L$ is a model, $x \in L$ and $\alpha < \kappa$,
and let $X_i\subseteq (x]_\alpha$ for all $i \in I$. 
Then $\bigcup_{i \in I} X_i$ and $\{\bigsqcup_\alpha X_i: i \in I\}$ are 
subsets of $(x]_\alpha$ and thus of $(x]_\beta$ for all 
$\beta < \alpha$. The following associativity property holds
for all $\beta \leq \alpha$: 
\begin{eqnarray}
\label{eq-l1}
\bigsqcup_\beta \{\bigsqcup _\alpha X_i : i \in I\} 
&=& 
\bigsqcup_{\beta} \bigcup_{i \in I} X_i.
\end{eqnarray}
\end{lem}

{\sl Proof.} 
When $\beta = \alpha$ this is due to the fact that for all $z \in L$, $\{\bigsqcup_\alpha X_i : i \in I\} \sqsubseteq_\alpha z$ iff\ $\bigcup_{i \in I} X_i \sqsubseteq_\alpha z$. Note that when $I$ is empty, then  both 
sides of (\ref{eq-l1}) are equal to the 
$\leq$-least element of $(x]_\alpha$.

Suppose now that $\beta < \alpha$.  Then $x =_\beta y$ holds for all $y \in \bigcup_{i \in I}X_i$,
moreover, $x =_\beta \bigsqcup_\alpha X_i$ for all $i \in I$. 
Hence, when $I$ is not empty,
 both sides of (\ref{eq-l1}) are equal to the $\leq$-least element of $[x]_\beta$.  
Otherwise, if $I$ is empty, both sides are equal to the $\leq$-least element of $(x]_\beta$. 
\eop

In \cite{ERfp}, it is proved that for any sequence $(x_\alpha)_{\alpha < \kappa}$ 
in a model $L$, there exists some $x \in L$ with $x_\alpha = x|_\alpha$ 
for all $\alpha < \kappa$ iff 
\begin{itemize}
\item $x_\alpha =_\alpha x_\beta$ for all $\alpha < \beta < \kappa$, and 
\item $x_\alpha$ is the $\leq$-least element of $[x_\alpha]_\alpha$, 
     for all $\alpha < \kappa$.
\end{itemize}
It follows that $x_\alpha \leq x_\beta$ for all $\alpha \leq \beta <\kappa$,
and if $\alpha + 1 < \kappa$, then $x_\alpha$ is a $\sqsubseteq_{\alpha+1}$-least 
element of $[x_\alpha]_\alpha$. Such sequences are called \emph{compatible}.
The element $x$ is uniquely determined by the compatible sequence $(x_\alpha)_{\alpha < \kappa}$.
It is given by $x = \bigvee_{\alpha < \kappa} x_\alpha$. Indeed, if $\alpha < \kappa$, 
then $x = \bigvee_{\alpha \leq \gamma < \kappa} x_\gamma =_\alpha x_\alpha$ by $Ax4$, 
since $x_\alpha =_\alpha x_\gamma$ for all $\alpha \leq \gamma < \kappa$. 
Hence $x|_\alpha = x_\alpha|_\alpha = x_\alpha$ for all $\alpha < \kappa$.

For later use we prove:

\begin{lem}
\label{lem-limitord}
Suppose that the sequence $(x_\gamma)_{\gamma < \kappa}$ is compatible
and $x = \bigvee_{\gamma < \kappa} x_\gamma$, so that $x =_\gamma x_\gamma$ for all $\gamma < \kappa$. 
Then for each limit ordinal $\alpha < \kappa$, $y = \bigvee_{\gamma < \alpha} x_\gamma$
is both the $\leq$-least and a $\sqsubseteq_\alpha$-least element of the set 
$(x]_\alpha = (x_\alpha]_\alpha$.  
\end{lem}

{\sl Proof.} 
Since the sequence $(x_\gamma)_{\gamma < \kappa}$ is compatible,
it is increasing w.r.t. $\leq$.  Let $\alpha < \kappa$ be a limit 
ordinal. Then for each $\beta < \alpha$, $y = \bigvee_{\gamma < \alpha} x_\gamma = 
\bigvee_{\beta \leq \gamma < \alpha} x_\gamma$. By compatibility, 
also $x_\beta = x_\gamma$ whenever $\beta \leq \gamma < \alpha$, 
hence by $Ax4$, $y =_\beta x_\beta$ for all $\beta < \alpha$, 
proving $y \in (x]_\alpha$. If $z \in (x]_\alpha$, then $z \in [x_\gamma]_\gamma = [x]_\gamma$ 
for all $\gamma < \alpha$. But each $x_\gamma$ is $\leq$-least in $[x_\gamma]_\gamma$, 
hence $x_\gamma \leq z$. Since this holds for all $\gamma$, $y = \bigvee_{\gamma < \alpha}x_\gamma \leq z$.
Thus, $y$ is the $\leq$-least element of $(x]_\alpha$, hence also 
a $\sqsubseteq_\alpha$-least element of $(x]_\alpha$
(cf. Axiom 3 in the case when $X$ is empty). \eop

For all $x,y\in L$ and $\alpha < \kappa$, let us write $x \sqsubset_\alpha y$ to denote that $x \sqsubseteq_\alpha y$ and $x \neq y$.
Moreover, we define $x \sqsubset y$ iff there is some $\alpha$ with 
$x \sqsubset_\alpha y$, and let $x \sqsubseteq y$ iff $x \sqsubset y$ or $x = y$.

 The following result was proved in \cite{ERfp}.

\begin{thm}
\label{thm-lattice}
Suppose that $L$ is a model. Then $(L, \sqsubseteq)$ is a complete 
lattice.
\end{thm}

\begin{remark}
Suppose that $L$ is a model. As usual, let $\bot$ denote the least element of $L$ 
w.r.t. $\leq$. Then $\bot$ is also the least element of $L$ w.r.t. $\sqsubseteq$. 
Indeed, for each $\alpha < \kappa$, $\bot \in (\bot]_\alpha$ and if $y \in (\bot]_\alpha$
then $\bot \leq y$ (as this inequality holds for all $y$). Hence, $\bigsqcup_\alpha \{\emptyset\} = \bot$ 
with $\emptyset$ considered as a subset of $(\bot]_\alpha$. Suppose now that $x \in L$,
$x \neq \bot$. Then there is a least ordinal $\alpha< \kappa $ with $\bot \not =_\alpha x$.
Then $x \in (\bot]_\alpha$, hence $\bot \sqsubseteq_\alpha x$, and since 
$\bot \not =_\alpha x$, we have $\bot \sqsubset_\alpha x$. 
Thus, for each $x \in L$, either $x = \bot$ or $\bot \sqsubset_\alpha x$ for some 
$\alpha < \kappa$. We conclude that $\bot \sqsubset x$. (It follows that 
$\bot \sqsubseteq_0 x$ for all $x$.) 
\end{remark}

\begin{expl}
In the standard model $V^Z$ defined in Example~\ref{expl-VZ}, the greatest element 
w.r.t. $\sqsubseteq$ is the function mapping each $z \in Z$ to $T_0$, which is 
the also the greatest element with respect to $\leq$. However, the greatest elements
with respect to the orderings need not be the same, not even in finite models.
Consider the lattice $L = \{0,1\}^2$ 
ordered by the relation $\leq$ as usual. Let $\kappa = 2$
and define $x_1x_2 \sqsubseteq_0 y_1y_2$ iff $x_1 = y_1$ or $x_1 \leq y_1$. 
Moreover, let $x_1x_2 \sqsubseteq_1 y_1y_2$ iff $x_1 = y_1 = 0$ and $x_2 \leq y_2$
or $x_1 = y_1 = 1$ and $x_2 \geq y_2$. Then $L$ is a model and the greatest elements of $L$ 
w.r.t. $\leq$ and $\sqsubseteq$ are $11$ and $10$, respectively.
\end{expl}

We now define \emph{$\alpha$-monotonic} and \emph{$\alpha$-continuous} functions, where $\alpha < \kappa$.
Suppose that $L$ and $L'$ are models. We say that a 
function $f: L \to L'$ is $\alpha$-monotonic if it preserves the preordering $\sqsubseteq_\alpha$,
i.e., when $x \sqsubseteq_\alpha y$ implies $fx \sqsubseteq_\alpha fy$ for all $x,y\in L$.
Moreover, we say that $f$ is $\alpha$-continuous if it is $\alpha$-monotonic 
and for all nonempty linearly ordered sets $(I,\leq)$
and $x_i\in L$ for $i \in I$, if $x_i \sqsubseteq_\alpha x_j$ for all $i \leq j$ in $I$,
then $$f(\bigsqcup_\alpha \{x_i :i \in I\}) =_\alpha \bigsqcup_\alpha \{fx_i : i \in I\},$$
or equivalently,  
$$
(f(\bigsqcup_\alpha \{x_i :i \in I\}))|_\alpha  = 
\bigsqcup_\alpha \{fx_i : i \in I\}. 
$$
(Note that since $x_i =_\beta x_j$ for all $\beta < \alpha$ and $i,j \in I$
and $f$ is $\alpha$-monotonic, also $fx_i =_\beta fx_j$ for all $\beta < \alpha$ and $i,j \in I$,
hence $\bigsqcup_\alpha \{x_i :i \in I\}$ and $\bigsqcup_\alpha \{fx_i :i \in I\}$ exist.)

\begin{expl}
\label{expl-complete lattices2}
Suppose that $L$ is a complete lattice viewed as a model
as in Example~\ref{expl-complete lattices}. Let $ \alpha < \kappa$. 
Then a function $f: L \to L$ is $0$-monotonic iff 
it is $\alpha$-monotonic
for all $\alpha < \kappa$ iff $f$ is monotonic with respect to $\leq$, 
and $f$ is $0$-continuous iff $\alpha$-continuous for all $\alpha < \kappa$
iff $f$ is continuous with respect to $\leq$, or simply just continuous:
 $f (\bigvee X) = \bigvee f(X)$ for all nonempty linearly ordered 
sets $X\subseteq L$ w.r.t. $\leq$.  
\end{expl} 

\begin{remark}
 A function between models that is $\alpha$-monotonic or $\alpha$-continuous 
for all $\alpha < \kappa$ is not necessarily monotonic w.r.t. the relation $\sqsubseteq$.
See \cite{ERfp}.
\end{remark}

We will make use of the next lemma without explicitly mentioning it. 

\begin{lem}
Suppose that $L,L',L''$ are models and $f: L \to L'$ 
and $g: L' \to L''$ are $\alpha$-monotonic (resp. $\alpha$-continuous),
where $\alpha$ is an ordinal less than $\kappa$. 
Then the function $g \circ f: L \to L''$ is 
also $\alpha$-monotonic (resp. $\alpha$-continuous). Moreover,
the identity function $\id_L: L \to L$ is $\alpha$-continuous
as is any constant function $L \to L'$. 
\end{lem}

Let $\Mod_m$ (resp. $\Mod_c$) denote the category of models 
and those functions between them which are $\alpha$-monotonic 
(resp. $\alpha$-continuous) for all $\alpha < \kappa$.

\begin{thm}
\label{thm-cc}
The categories $\Mod_m$ and $\Mod_c$ are cc's. 
\end{thm}

{\sl Proof.} We need to show that $\Mod_m$ (resp. $\Mod_c$) has a terminal object and 
binary products. In fact, it is easy to see
that $\Mod_m$ (resp. $\Mod_c$) has all products and that products can be constructed 
pointwise. Thus, if 
$L_i = (L_i,\leq_i,(\sqsubseteq_{i,\alpha})_{\alpha < \kappa})$ 
is a model for each $i \in I$, 
where $I$ is any set, then the cartesian product $L = \prod_{i \in I} L_i$,
equipped with the pointwise order relations $\leq$ and $\sqsubseteq_\alpha$,
$\alpha < \kappa$, defined for $x = (x_i)_{i\in I}$ and $y = (y_i)_{i \in I}$ in $L$ 
by $x \leq y$ iff $x_i \leq_i y_i$ for all $i \in I$ 
and $x\sqsubseteq_\alpha y$ iff $x_i \sqsubseteq_{i,\alpha} y_i$ for all $i \in I$, 
is also a model. It follows that $\bigvee X$ can be computed pointwise for all $X \subseteq L$. 
A similar fact is true for $\bigsqcup_\alpha X$ whenever $X \subseteq (x]_\alpha$
for some $x \in L$ and $\alpha < \kappa$.

In both categories,
the  projections $\pi_{L_j}^{\prod_{i \in I} L_i}:  \prod_{i \in I} L_i \to L_j$, for $j \in I$, 
are the usual projection functions. 
\eop 

\begin{remark}
\label{rem-prod}
Suppose that $L_i$ is a model as above for each $i \in I$. 
Let $x = (x_i)_{i\in I}$ and $y = (y_i)_{i \in I}$ in the 
product model $L = \prod_{i \in I} L_i$. Then $x \sqsubseteq y$ 
in $L$ iff either $x = y$, or there is some $\alpha < \kappa$
such that for all $i\in I$, $x_i \sqsubseteq_{i,\alpha} y_i$,
moreover, there is some $j \in I$ with $x_j \sqsubset_{j,\alpha} y_j$. 
\end{remark}

Below we will often make use of the following simple fact. 

\begin{lem}
Suppose that $L,L',L''$ are models and $\alpha < \kappa$. Then a function $f: L' \times L'' \to L$ 
is $\alpha$-monotonic (resp. $\alpha$-continuous) iff the following conditions hold:
\begin{itemize}
\item
For each fixed $x \in L'$, the function $_xf: L'' \to L$ 
defined by $_xf y = f(x,y)$ is $\alpha$-monotonic ($\alpha$-continuous). 
\item
For each fixed $y \in L''$, the function $f_y: L' \to L$ 
defined by $f_y x = f(x,y)$ is $\alpha$-monotonic ($\alpha$-continuous). 
\end{itemize}
\end{lem}

\begin{expl}
\label{expl-VZ2} \cite{ERfp,RW}
Consider the linearly ordered complete lattice $V$ of truth values of Example~\ref{expl-VZ}
and let $Z$ be a set. 
Define $\vee$ and $\wedge$ as the binary supremum and infimum operations on $V$, 
and define $\neg: V \to  V$ by $\neg F_\alpha = T_{\alpha + 1}$, 
$\neg T_\alpha = F_{\alpha + 1}$, for all $\alpha < \Omega$,
 and $\neg 0 = 0$. Extend these 
operations to $V^Z$ pointwise, so that $(f\vee g)z = fz \vee gz$ 
for all $z \in Z$, etc. Then $\vee,\wedge,\neg$ are $\alpha$-continuous
functions over $V^Z$ for all $\alpha < \Omega$.
\end{expl}

\section{Stratified least fixed points}
\label{sec-fp}

In this section we recall a fixed point theorem (Theorem~\ref{thm-fp}) from \cite{ERfp}
involving those functions over a model $L$ which are $\alpha$-monotonic 
for all $\alpha < \kappa$. (Recall that $\kappa$ is a fixed nonzero ordinal.)
Then we extend this operation to a \emph{parametrized fixed point operation}
$$f : L \times L' \quad \mapsto \quad f^\dagger: L' \to L,$$
where $L$ and $L'$ are models and $f$ is $\alpha$-monotonic for all $\alpha < \kappa$, 
and prove that $f^\dagger$ is also $\alpha$-monotonic for all $\alpha < \kappa$. 
Moreover, we prove that when $f$ is $\alpha$-continuous for all $\alpha < \kappa$, 
then so is $f^\dagger$.

\begin{thm}
\label{thm-fp}
Suppose that $L$ is a model and 
$f: L \to L$ is $\alpha$-monotonic for all $\alpha < \kappa$.
Then $f$ has a least pre-fixed point with respect to 
the partial order $\sqsubseteq$ which is also a fixed point.
\end{thm}

Thus, the theorem asserts that there is some $x\in L$ 
with $fx \sqsubseteq x$ and such that for all $y\in L$,
if $fy \sqsubseteq y$ then $x \sqsubseteq y$.
Moreover, $x$ is a fixed point, i.e., $fx = x$. In particular,
$x$ is the unique least fixed point of $f$ w.r.t. the 
order relation $\sqsubseteq$.
The proof of Theorem~\ref{thm-fp} in \cite{ERfp} provides a  
construction of the least fixed point by a two level transfinite 
sequence of approximations. We will describe the construction in more 
detail below. 
Since every $\alpha$-continuous function is $\alpha$-monotonic,
the theorem also applies to functions that are $\alpha$-continuous
for all $\alpha <\kappa$. For such functions, the inner level of 
the transfinite sequence of approximations terminates in $\omega$ 
steps, where $\omega$ denotes the first infinite ordinal.

We will be concerned with parametrized fixed points. 
Suppose that $L,L'$ are models and 
$f: L \times L' \to L$ is $\alpha$-monotonic for all $\alpha < \kappa$.
Then for each fixed $y\in L'$, the function $f_y: L \to L$ 
defined by $f_y x = f(x,y)$ is also $\alpha$-monotonic for all 
$\alpha < \kappa$ and thus 
by Theorem~\ref{thm-fp} has a least (pre-)fixed point with respect 
to the ordering $\sqsubseteq$.
Let us denote this least (pre-)fixed point by $f^\dagger y$. 
Then $f^\dagger$, as a function of $x$, maps $L'$ into $L$. 

\begin{expl}
\label{expl-complete lattices3}
As explained in Example~\ref{expl-complete lattices},
each complete lattice $L = (L,\leq)$ gives rise to a model 
for any nonzero ordinal $\kappa$. Moreover,
by Example~\ref{expl-complete lattices2},
 a function $f : L \to L'$ between complete lattices $L$ and $L'$
is $\alpha$-monotonic for all $\alpha < \kappa$ iff it is monotonic
with respect to the ordering $\leq$.
Similarly, $f$ is $\alpha$-continuous for all $\alpha < \kappa$ iff 
it is continuous w.r.t. $\leq$. 
Thus, in this case Theorem~\ref{thm-fp} asserts that for a complete 
lattice $(L,\leq)$, every monotonic function $L \to L$ 
has a least pre-fixed point (w.r.t. $\leq$)  which is a fixed point.
This is a part of the Knaster-Tarski fixed point 
theorem, see \cite{Daveyetal,Tarski}. 
In particular, for all complete lattices $L,L'$ and monotonic 
functions $f: L \times L' \to L$ and $y\in L'$, $f^\dagger y $ is 
the least (pre-)fixed point of the function $f_y : L \to L$
mapping each $x \in L$ to $f(x,y)$. 
\end{expl}

\begin{expl} 
\label{expl-logic program}\cite{ERfp,RW}
Suppose that $Z$ is a denumerable set of propositional variables and $P$
is an at most countably infinite propositional logic program over $Z$, 
possibly involving negation. Thus $P$ 
is a countable set of instructions of the form 
$z \leftarrow \ell_1 \wedge \cdots \wedge \ell_k$, where $z \in Z$ and $\ell _i$ 
is a literal for each $i$. Consider the model $L = V^Z$ of `interpretations' defined earlier,
cf. Example~\ref{expl-VZ}.   
 Then $P$ induces a function $f_P: L \to L$
which maps an interpretation $I \in L$ to the 
interpretation $J = f_P(I)$ such that 
$J(z) = \bigvee_{z  \leftarrow \ell_1\wedge \cdots \wedge \ell_k \in P} (I(\ell_1) \wedge \cdots \wedge I(\ell_k))$,
where for a negative literal $\ell = \neg y$ and $\alpha < \Omega$,  $I(\ell) = T_{\alpha+1}$ if $I(y) = F_\alpha$,
$I(\ell) = F_{\alpha+1}$ if $I(y) = T_\alpha$, and $I(\ell) = 0$ if $I(y) = 0$. 
Then $f_P$ is $\alpha$-monotonic for all $\alpha < \Omega$. The semantics of $P$ 
is defined in \cite{RW} as the least fixed point of $f_P$ w.r.t. $\sqsubseteq$. 

For example, consider the program $P$: 
\begin{eqnarray*}
p &\leftarrow & \neg q\\
q &\leftarrow & \neg r\\
s &\leftarrow & p\\
s &\leftarrow & \neg s\\
t &\leftarrow & 
\end{eqnarray*}
Then the least fixed point of $f_P$ w.r.t. $\sqsubseteq$ is: $(r,F_0), (q,T_1), (p,F_2), (s,0), (t, T_0)$. 
Intuitively $q$ is `less true' than $t$, since $q$ is true only because $r$ is false by default, while 
there is an instruction declaring $t$ to be true. This is reflected by the least fixed point.  
\end{expl}

The construction of $f^\dagger y$ mentioned above makes 
use of the following lemma from \cite{ERfp}, slightly 
adjusted to the parametrized setting.

\begin{lem}
\label{lem-falpha}
Suppose that $L,L'$ are models and $f: L \times L' \to L$ is $\alpha$-monotonic,
where $\alpha < \kappa$. If $x\in L$, $y \in L'$ and $\alpha < \kappa$
with $x \sqsubseteq_\alpha f(x,y)$, then there is some $z \in L$ with the 
following properties:
\begin{itemize}
\item $x \sqsubseteq_\alpha z =_\alpha f (z,y)$,
\item if $z' \in L$ with $x \sqsubseteq_\alpha z'$ and $f(z',y) \sqsubseteq_\alpha z'$, 
then $z \sqsubseteq_\alpha z'$,
\item $z$ is the $\leq$-least element of the set $[z]_\alpha$, and if 
 $\alpha+1 < \kappa$, then $z$ is also a $\sqsubseteq_{\alpha + 1}$-least element of $[z]_\alpha$.
\end{itemize} 
\end{lem} 
It follows that $z$ is uniquely determined as a function of $x$ and $y$
and we denote it by $f_\alpha(x,y)$.
Indeed, if $z$ and $z'$ both satisfy the above conditions, then 
$z\sqsubseteq_\alpha z'$ and $z' \sqsubseteq_\alpha z$, hence $z =_\alpha z'$ and 
$[z]_\alpha = [z']_\alpha$, so that $z$ and $z'$ are the $\leq$-least elements of 
the same set. Moreover, if $\alpha + 1 < \kappa$, then $z \sqsubseteq_{\alpha+1} f(z,y)$, since $z$ is a $\sqsubseteq_{\alpha + 1}$-least element of $[z]_\alpha
= [f(z,y)]_\alpha$.

The element $z = f_\alpha(x,y)$ can be constructed by approximating it with the 
following sequence $(x_\gamma)_\gamma$, where $\gamma$ ranges over the ordinals. 
Let $x_0 = x$ and $x_\gamma = f(x_\delta,y)$ when  $\gamma = \delta + 1$ is 
a successor ordinal. When $\gamma$ is a limit ordinal, define 
$x_\gamma = \bigsqcup_\alpha \{x_\delta : \delta < \gamma\}$. 
Then $x_\beta \sqsubseteq_\alpha x_\gamma$ for all ordinals $\beta$ and $\gamma$ with 
$\beta < \gamma$. Thus there is a least ordinal $\lambda_0$ with $x_{\lambda_0} =_\alpha 
x_{\lambda_0 + 1}$. It follows that $x_\beta =_\alpha x_\gamma$ for all 
$\beta$ and $\gamma$ with $\lambda_0 \leq \beta < \gamma$. The element $z$ 
is $x_{\lambda}$ for the least limit ordinal $\lambda$ with $\lambda_0 \leq \lambda$. 
In the case when $f$ is $\gamma$-continuous for all $\gamma < \kappa$,
the ordinal $\lambda$ is $\omega$, so that the construction stops in $\omega$
steps. 

Now $z = f^\dagger y$ can be constructed as follows. For each $\alpha < \kappa$, let 
$x_\alpha = \bigvee_{\beta < \alpha} z_\beta$ and $z_\alpha = f_\alpha(x_\alpha,y)$,
so that $x_0 = \bot$. This construction is legitimate, since as shown in \cite{ERfp},
$x_\alpha \sqsubseteq_\alpha f(x_\alpha,y)$ for all $\alpha < \kappa$.
Moreover, the sequence 
$(z_\alpha)_{\alpha < \kappa}$ is compatible and $z = \bigvee_{\alpha < \kappa} z_\alpha$,
so that $z|_\alpha = z_\alpha$ for all $\alpha < \kappa$.

\begin{remark}
Sometimes we will apply the dagger operation to functions $f: L \to L$, where $L$ 
is a model and $f$ is $\alpha$-monotonic for all $\alpha < \kappa$. In this case 
we identify $L$ with $L \times \one$, where $\one$ is a fixed one-element model,
so that $f^\dagger : \one \to L$, which is in turn conveniently identified with an element of $L$. 
\end{remark} 

We will make use of the following lemmas concerning the functions $f_\alpha$. 

\begin{lem}
\label{lem-mon}
Suppose that $L,L'$ are models
and $f:L \times L' \to L$ is $\alpha$-monotonic,
where $\alpha <\kappa$. Suppose that $x,x'\in L$ and $y,y'\in L'$ 
with $x \sqsubseteq_\alpha x'$ and $y \sqsubseteq_\alpha y'$,
moreover, $x \sqsubseteq_\alpha f(x,y)$ and $x' \sqsubseteq_\alpha f(x',y')$.  
Let $z = f_\alpha(x,y)$ 
and $z' = f_\alpha (x',y')$. Then $z \sqsubseteq_\alpha z'$. And if
$x=_\alpha x'$ and $y =_\alpha y'$ then $z =_\alpha z'$, and in fact 
$z = z'$.
\end{lem}

{\sl Proof.} 
First note that since $x \sqsubseteq_\alpha f(x,y)$ and $x' \sqsubseteq_\alpha f(x',y')$,
both $z = f_\alpha(x,y)$  and $z' = f_\alpha (x',y')$ exist. 
Since $f$ is $\alpha$-monotonic, we have 
$f(z',y)\sqsubseteq_\alpha f(z',y') \sqsubseteq_\alpha z'$.
Also, $x\sqsubseteq_\alpha x' \sqsubseteq_\alpha z'$. 
It follows by the 2nd clause of Lemma~\ref{lem-falpha} that $z \sqsubseteq_\alpha z'$.
Suppose now that $x=_\alpha x'$ and $y =_\alpha y'$.
Then $z \sqsubseteq_\alpha z'$ and $z' \sqsubseteq_\alpha z$,
thus $z =_\alpha z'$. Since $z$ is the $\leq$-least element
of $[z]_\alpha$ and  $z'$ is the $\leq$-least element
of $[z']_\alpha$, and since $[z]_\alpha = [z']_\alpha$,
it follows that $z = z'$.
\eop

\begin{lem}
\label{lem-cont}
Suppose that $L,L'$ are models
and $f:L \times L' \to L$ is $\alpha$-continuous,
where $\alpha <\kappa$. Suppose that $(I,\leq)$ 
is a nonempty linearly ordered set and $x_i\in L$, $y_i\in L'$ for all $i \in I$
such that $x_i \sqsubseteq_\alpha x_j$ and
$y_i \sqsubseteq_\alpha y_j$ whenever $i \leq j$ in $I$,
moreover, 
$x_i \sqsubseteq_\alpha f(x_i,y_i)$ for all $i \in I$. 
Then
\begin{eqnarray*}
\bigsqcup_\alpha \{x_i : i\in I\}
&\sqsubseteq_\alpha&
f (\bigsqcup_\alpha\{x_i : i\in I\}, \bigsqcup_\alpha\{y_i : i\in I\}) 
\end{eqnarray*} 
and
\begin{eqnarray*}
f_\alpha (\bigsqcup_\alpha\{x_i : i\in I\}, \bigsqcup_\alpha\{y_i : i\in I\}) 
&=&
\bigsqcup_\alpha \{f_\alpha(x_i,y_i) : i \in I\}.
\end{eqnarray*}
\end{lem}

{\sl Proof.}
First, since $x_i \sqsubseteq_\alpha f(x_i,y_i)$ for all $i \in I$ 
and $f$ is $\alpha$-continuous,  
\begin{eqnarray*}
\bigsqcup_\alpha \{x_i : i\in I\}
&\sqsubseteq_\alpha& 
\bigsqcup_\alpha \{f(x_i,y_i) : i \in I\}\\
&=_\alpha &
f (\bigsqcup_\alpha\{x_i : i\in I\}, \bigsqcup_\alpha\{y_i : i\in I\}).
\end{eqnarray*} 
Thus, $f_\alpha (\bigsqcup_\alpha\{x_i : i\in I\}, \bigsqcup_\alpha\{y_i : i\in I\})$
exists.

Define $f^n : L \times L' \to L$ for $n \geq 0$ by $f^0 = \pi^{L \times L'}_{L}$ 
and $f^n = f \circ \langle f^{n-1}, \pi^{L \times L'}_{L'}\rangle$ for $n > 0$.
Thus, for all $x \in L$ and $y \in L'$, $f^0(x,y) = x$ and $f^n(x,y) = 
f(f^{n-1}(x,y), y)$, for all $n > 0$. Since $f$ is $\alpha$-continuous, so is 
$f^n$ for all $n \geq 0$. It follows from our assumptions
that $f^n(x_i,y_i) =_\beta f^m(x_j,y_j)$ for all $n,m\geq 0$, $i,j\in I$
and $\beta < \alpha$. Moreover, using $\alpha$-continuity, 
\begin{eqnarray*}
f_\alpha (\bigsqcup_\alpha\{x_i : i\in I\}, \bigsqcup_\alpha\{y_i : i\in I\}) 
&=&
\bigsqcup_\alpha \{f^n (\bigsqcup_\alpha\{x_i : i\in I\}, \bigsqcup_\alpha\{y_i : i\in I\}) : n \geq 0\}\\
&=_\alpha& 
\bigsqcup_\alpha \{f^n(x_i,y_i) : i\in I,\  n \geq 0\}.
\end{eqnarray*}
However, by Lemma~\ref{lem-falpha}, $f_\alpha (\bigsqcup_\alpha\{x_i : i\in I\}, \bigsqcup_\alpha\{y_i : i\in I\})$
is the $\leq$-least element of $[f_\alpha (\bigsqcup_\alpha\{x_i : i\in I\}, \bigsqcup_\alpha\{y_i : i\in I\})]_\alpha$ and by definition, 
$\bigsqcup_\alpha \{f^n(x_i,y_i) : i\in I,\  n \geq 0\}$ is the $\leq$-least element of $[\bigsqcup_\alpha \{f^n(x_i,y_i) : i\in I,\  n \geq 0\}]_\alpha$, hence 
\begin{eqnarray*}
f_\alpha (\bigsqcup_\alpha\{x_i : i\in I\}, \bigsqcup_\alpha\{y_i : i\in I\}) 
&=&\bigsqcup_\alpha \{f^n(x_i,y_i) : i\in I,\  n \geq 0\}.
\end{eqnarray*}

Also,
\begin{eqnarray*}
\bigsqcup_\alpha \{f_\alpha(x_i,y_i) : i \in I\}
&=&
\bigsqcup_\alpha \{ \bigsqcup_\alpha \{ f^n(x_i,y_i) : n \geq 0\} : i \in I\}\\
&=& 
\bigsqcup_\alpha \{f^n(x_i,y_i) : i\in I,\  n \geq 0\},
\end{eqnarray*}
by Lemma~\ref{lem-assoc}. 
\eop 

Now we can prove that if $f$ is $\alpha$-monotonic or $\alpha$-continuous 
for all $\alpha < \kappa$, then so is $f^\dagger$. 

\begin{prop}
Suppose that $L,L'$ are models and 
$f: L \times L' \to L$ is $\alpha$-monotonic for all $\alpha < \kappa$.
Then the function $f^\dagger: L'\to L$ is also $\alpha$-monotonic
for all $\alpha < \kappa$. And if $f$ is $\alpha$-continuous 
for all $\alpha < \kappa$, then the same holds for $f^\dagger$.
\end{prop}

{\sl Proof.} 
Suppose first that $f$ is $\alpha$-monotonic for all $\alpha < \kappa$.
Let $y \sqsubseteq_\alpha y'$ in $L'$,
where $\alpha < \kappa$, and denote $z = f^\dagger y$ and $z' = f^\dagger y'$.
We want to show that $z \sqsubseteq_\alpha z'$.

For each $\gamma < \kappa$, let $x_\gamma = \bigvee_{\delta < \gamma} z_\delta$
and $z_\gamma = f_\gamma (x_\gamma,y)$. Symmetrically, let 
$x_\gamma' = \bigvee_{\delta < \gamma} z_\delta'$ and 
$z_\gamma' = f_\gamma (x_\gamma',y')$. Thus $x_0 = x_0' = \bot$.
We know that $x_\gamma \sqsubseteq_\gamma f(x_\gamma,y)$ 
and $x'_\gamma \sqsubseteq_\gamma f(x'_\gamma,y')$ 
for all $\gamma < \kappa$. Moreover, the sequences $(z_\gamma)_{\gamma < \kappa}$ 
and $(z'_\gamma)_{\gamma < \kappa}$ are compatible and 
$z_\gamma = z|_\gamma$ and $z'_\gamma = z'|_\gamma$ 
for all $\gamma < \kappa$,
and $z = \bigvee_{\gamma < \kappa} z_\gamma$ and $z' = \bigvee_{\gamma < \kappa} z_\gamma'$.
Thus, $z =_\alpha z_\alpha$ and $z' =_\alpha z_\alpha'$, so that
$z \sqsubseteq_\alpha z'$ holds exactly when $z_\alpha \sqsubseteq_\alpha z'_\alpha$.

It follows by induction on $\beta$ using Lemma~\ref{lem-mon} that $z_\beta = z_\beta'$ and $x_\beta = x_\beta'$
for all $\beta < \alpha$. Indeed,  $x_0 = x_0' = \bot$ and if $\alpha > 0$ then $z_0 = z_0'$ 
by Lemma~\ref{lem-mon} since $y =_0 y'$. 
And if $0 < \beta < \alpha$ and the claim holds for all $\gamma < \beta$, then 
$x_\beta = \bigvee_{\gamma < \beta} z_\gamma = \bigvee_{\gamma < \beta}z_\gamma' = x'_\beta$,
and then $z_\beta = z'_\beta$ by Lemma~\ref{lem-mon} since $y =_\beta y'$. 
Also, $x_\alpha = \bigvee_{\beta < \alpha} z_\beta 
= \bigvee_{\beta < \alpha}z_\beta'  = x_\alpha'$. Since $y \sqsubseteq_\alpha y'$,
it follows now that $z_\alpha = f_\alpha(x_\alpha ,y) \sqsubseteq_\alpha f_\alpha (x_\alpha', y') = z_\alpha'$.

Suppose next that $f$ is $\alpha$-continuous for all $\alpha < \kappa$.
We prove that $f^\dagger$ is also $\alpha$-continuous for all $\alpha < \kappa$. 
To this end, let $\alpha < \kappa$, $(I,\leq)$ be a nonempty linearly ordered set 
and $y_i\in L$ for all $i \in I$ such that
$y_i \sqsubseteq_\alpha y_j$ whenever $i,j\in I$ with $i \leq j$. Let $y = \bigsqcup_\alpha \{y_i :i\in I\}$.
For each $i\in I$ and $\gamma < \kappa$, define $x_{i,\gamma} = \bigvee_{\beta < \gamma} z_{i,\beta}$
and $z_{i,\gamma} = f_\gamma (x_{i,\gamma},y_i)$. Moreover,
define $x_\gamma = \bigvee_{\beta < \gamma} z_\beta$
and $z_\gamma = f_\gamma (x_\gamma,y)$. 
We know that $x_{i,\gamma} \sqsubseteq_\gamma f(x_{i,\gamma},y_i)$ for all $i \in I$ and $\gamma < \kappa$.
Similarly, $x_\gamma \sqsubseteq_\gamma f(x_\gamma,y)$ for all $\gamma < \kappa$.

Let $z = f^\dagger y$ and $z_i = f^\dagger y_i$ for all $i \in I$.
We already know that $z = \bigvee_{\gamma < \kappa} z_\gamma$ and
$z_i = \bigvee_{\gamma < \kappa} z_{i,\gamma}$ for all $i \in I$.  
Also, $z_i \sqsubseteq_\alpha z_j \sqsubseteq_\alpha z$ 
for all $i,j \in I$ with $i \leq j$. 
We want to prove that $z =_\alpha \bigsqcup_\alpha \{ z_i : i \in I\}$.
Since $z =_\alpha  z_\alpha$ and $z_i =_\alpha z_{i,\alpha}$ 
for all $i \in I$, this holds if $z_\alpha =_\alpha \bigsqcup_\alpha \{z_{i,\alpha} : i \in I\}$.

It follows by induction using Lemma~\ref{lem-mon} that 
$x_\gamma = x_{i,\gamma}$ and $z_i = z_{i,\gamma}$ for all 
$i \in I$ and $\gamma < \alpha$. Indeed, suppose that 
$\gamma < \alpha$, $x_\delta = x_{i,\delta}$ and $z_\delta = z_{i,\delta}$ for all 
$i \in I$ and $\delta < \gamma$. Then $x_\gamma = \bigvee_{\delta < \gamma} x_\delta 
= \bigvee_{\delta < \gamma} x_{i,\delta}  = x_{i,\gamma}$ for all $i \in I$.
Moreover, since $y =_\gamma y_i$ for all $i \in I$, by Lemma~\ref{lem-mon} also 
$z_\gamma = f_\gamma(x_\gamma,y) = f_\gamma(x_{i,\gamma},y_i) = z_{i,\gamma}$ 
for all $i \in I$. Similarly, $x_\alpha = x_{i,\alpha}$ for all $i \in I$. 
It follows eventually from Lemma~\ref{lem-cont} that 
\begin{eqnarray*}
z_\alpha 
&=& 
f_\alpha(x_\alpha, y) \\
&=& 
f_\alpha (\bigsqcup_\alpha \{x_{i,\alpha} : i \in I\},
\bigsqcup_\alpha \{y_i : i \in I\})\\
&=& 
\bigsqcup_\alpha \{f_\alpha(x_{i,\alpha},y_i) : i \in I\}\\
&=& 
\bigsqcup_\alpha \{z_{i,\alpha} : i \in I\}.
\end{eqnarray*}
\eop

\section{The cartesian fixed point identities} 
\label{sec-cartesian}

An  \emph{external dagger operation} \cite{BEccc} on a cartesian category assigns 
a morphism $f^\dagger : L' \to L$ to each morphism $f: L \times L' \to L$.
In particular, $\Mod_m$ and $\Mod_c$ are equipped with an external dagger 
operation. In this section, we prove that
with respect to the cartesian structure, the dagger operation on these categories 
satisfies the standard identities of fixed point operations
described by the axioms of iteration theories \cite{BEbook}.

We recall that a cartesian category is a category with finite products. 
We will assume that in each cartesian category, a terminal object 
$\one$ is fixed, and for each pair of objects $L,L'$, we assume a fixed 
product object $L \times L'$ and specified projection morphisms 
$\pi^{L \times L'}_L : L \times L' \to L$ 
and $\pi^{L \times L'}_{L'} : L \times L' \to L'$. 
Moreover, we assume that product is `associative on the nose',
so that in particular $L \times (L' \times L'') = (L \times L') \times L''$
and 
\begin{eqnarray*}
\pi^{L' \times L''}_ {L''} \circ \pi^{L \times(L' \times L'')}_{L' \times L''} 
&=& 
\pi ^{(L \times L')\times L''}_{L''},
\end{eqnarray*}
for all objects $L,L',L''$, etc. We identify an object 
$L \times \one$ with $L$ and a projection
$\pi^{L \times \one}_L$ with $\id_{L}$.

\emph{Some notation}. In any cartesian category, for any morphisms 
$f: L'' \to L$ and $g: L'' \to L'$ we denote by $\langle f, g \rangle$ 
the \emph{pairing} of $f$ and $g$, ie., the unique morphism $h: L'' \to L \times L'$ with 
$f = \pi^{L \times L'}_L \circ h$ and $f = \pi^{L \times L'}_{L'} \circ h$.
Note that in $\Mod_m$ or $\Mod_c$, $\langle f,g \rangle x = (fx,gx)$ 
for all $x \in L''$.
Moreover, for $f: L' \to L$ and $g: K' \to K$, we let 
$f \times g$ denote the morphism $\langle f \circ \pi^{L' \times K'}_{L'}, 
g \circ \pi^{L' \times K'}_{K'} \rangle : L' \times K' \to L \times K$. 
Thus, in $\Mod_m$ or $\Mod_c$, $(f \times g)(x,y) = (fx,gy)$
for all $(x,y) \in L' \times K'$. These operations are associative. 
We define the \emph{tupling} $\langle f_1,\ldots,f_n \rangle : L' \to 
\prod_{i = 1}^n L_i$ of morphisms $f_i : L' \to L_i$, $i = 1,\ldots,n$ 
by repeated applications of the pairing operation.

We now review one of the axiomatizations 
of iteration categories (or iteration theories) 
from \cite{BEbook,Esaxioms}. 
(Actually only cartesian categories generated by a single
object were treated in \cite{BEbook}, but the generalization 
is straightforward, see eg. \cite{BEccc,EsAC,EsMSCS}.)

\emph{Fixed point identity}
\begin{eqnarray*}
f^\dagger &=& f \circ \langle f^\dagger, \id_{L'} \rangle,\quad f: L \times L' \to L
\end{eqnarray*}

\emph{Parameter identity}
\begin{eqnarray*}
f^\dagger\circ g  &=& (f \circ (\id_L \times g))^\dagger,\quad f: L \times L' \to L,
\ g: L'' \to L'
\end{eqnarray*}

\emph{Composition identity}
\begin{eqnarray}
\label{eq-comp}
(g\circ \langle f, \pi_{L''}^{L \times L''}\rangle)^\dagger 
&=& 
g \circ \langle (f\circ \langle g, \pi_{L''}^{L' \times L''} \rangle)^\dagger,  \id_{L''}\rangle,
\end{eqnarray}
where $f: L \times L'' \to L'$ and $g: L' \times L'' \to L$.

\emph{Double dagger identity}
\begin{eqnarray*}
(f \circ ( \Delta_L \times \id_{L'}))^\dagger &=& f^{\dagger\dagger}, \quad f: L \times L \times L' \to L
\end{eqnarray*} 
(Here, $\Delta_L $ is the diagonal morphism $\langle \id_L,\id_L \rangle: L \to L \times L$.) 

\emph{Commutative identities}
\begin{eqnarray*}
\pi \circ \langle f \circ (\rho_1 \times \id_{L'}), \ldots, f \circ (\rho_n \times \id_{L'}) \rangle^\dagger
&=& (f \circ (\Delta_L^n \times \id_{L'}))^\dagger
\end{eqnarray*} 
where $n > 1$, $f: L^n \times L' \to L$, $\Delta_L^n =  \langle \id_L,\ldots,\id_L \rangle$ is the diagonal
morphism $L \to L^n$, the $\rho_i: A^n \to A^n$ are tuplings of projections,
and $\pi$ denotes the first projection $L^n \to L$.\footnote{Actually the above identities 
are an equivalent form of the original commutative identities \cite{Esaxioms}.}

Following \cite{BEbook}, we say that a cartesian category equipped with an external 
dagger operation is a \emph{Conway category} (or \emph{Conway theory}) if it satisfies the 
parameter, composition and double dagger identities. Moreover, we say that a cartesian 
category equipped with an external dagger operation is an \emph{iteration category} 
(or an \emph{iteration theory}) if it is a Conway category satisfying the commutative 
identities.

Before proceeding to prove that $\Mod_m$ and $\Mod_c$ are iteration categories,
we recall some facts from \cite{BEbook}. It is clear that 
the fixed point identity is an instance of the composition identity. 
Also, Conway categories satisfy several other well-known
identities including the pairing identity (or Beki\'c identity)
\cite{Bekic,ScottDeBakker}. 
It is immediately clear that in Conway categories, 
the commutative identities are implied 
by the following quasi-identity.

\emph{Weak functorial dagger}
\begin{eqnarray*}
f \circ (\Delta_L^n \times \id_{L'})= \Delta_L^n \circ g
&\Rightarrow&
f^\dagger = \Delta_L^n \circ g^\dagger,
\end{eqnarray*}
where $f: L^n \times L' \to L^n$, $g: L \times L' \to L$
and $\Delta_L^n$ denotes the diagonal morphism 
$L \to L^n$. 
For simplifications of the commutative identities, we refer to 
\cite{Esgroup,EsAC,EsMSCS}.

\begin{remark}
\label{rem-complete lattices4}
Consider the category $\mathcal{C}$ of complete lattices and 
monotonic or continuous functions. By Example~\ref{expl-complete lattices3}, 
the least fixed point operation is an external dagger operation on 
$\mathcal{C}$. It is known that each of the above identities 
as well as the weak functorial dagger implication holds in $\mathcal{C}$. 
Moreover, as shown in \cite{BEbook,Esaxioms}, an identity involving the 
cartesian operations and dagger holds in $\mathcal{C}$ 
iff it holds in all iteration categories.
For a generalization of this completeness result involving partially ordered sets 
and monotonic functions with enough least fixed points 
or least pre-fixed points, see \cite{Espark}.
For initial fixed points we refer to \cite{BEccc,EsikLabella}. 
\end{remark}

\begin{thm}
\label{thm-iteration cat}
$\Mod_m$ is an iteration category with a weak functorial dagger.  
\end{thm}

{\sl Proof.}
It is clear from the definition of dagger that the fixed point identity holds.
Due to the `pointwise' definition of dagger, the parameter 
identity also holds. Indeed, let $f: L \times L' \to L$ and $g: L'' \to L'$,
where $L,L',L''$ are models and $f,g$ are $\alpha$-monotonic for all 
$\alpha < \kappa$. We want to show that $(f\circ (\id_L \times g))^\dagger
= f^\dagger \circ g$. To this end, let $z \in L''$.
By definition, $(f\circ (\id_L \times g))^\dagger z$ is the $\sqsubseteq$-least 
$x \in L$ with $f(x,gz) = f(\id_L \times g)(x,z) \sqsubseteq x$.
Clearly, $f^\dagger gz $ is the same element.

To prove that the composition identity holds, suppose first that 
$L,L'$ are models and $f: L \to  L'$ and $g: L'  \to L$ 
are $\alpha$-monotonic for all $\alpha < \kappa$. 
Let $h = g\circ f : L \to L $ and $k = f \circ g: L' \to L'$. We want to show that 
$h^\dagger = g \circ k^\dagger$. Our argument uses the 
explicit construction of $h^\dagger$ and $k^\dagger$. 

Let $x_\alpha = \bigvee_{\beta < \alpha} y_\beta$ 
and $y_\alpha = h_\alpha x_\alpha$ for all $\alpha < \kappa$,
so that $x_0 = \bot$, the least element of $L$.
Similarly, let $x_\alpha' = \bigvee_{\beta < \alpha} y_\beta'$ 
and $y_\alpha' = k_\alpha x_\alpha'$ for all $\alpha < \kappa$.
Thus, $x_0' = \bot'$, the least element of $L'$. We know
that the sequences 
$(y_\alpha)_{\alpha< \kappa}$ and $(y_\alpha')_{\alpha < \kappa}$ 
are compatible, moreover,  $h^\dagger = y = \bigvee_{\alpha < \kappa}y_\alpha$,
$k^\dagger = y' = \bigvee_{\alpha < \kappa}y'_\alpha$. 
Also, $y|_\alpha = y_\alpha$ and  $y'|_\alpha = y'_\alpha$,
for all $\alpha < \kappa$. 

We show by induction on $\alpha$ that $y_\alpha =_\alpha g y_\alpha'$.
We will make use of the following lemma.

\begin{lem}
\label{lem-zz}
Suppose that $x\in L$ and  $x'\in L'$
with $x \sqsubseteq_\alpha gx'$ and $x' \sqsubseteq_\alpha fx$,
where $\alpha < \kappa$. 
Let $y = h_\alpha x$ and $y' = k_\alpha x'$. Then 
$y =_\alpha gy'$ and $y' =_\alpha  fy$. 
\end{lem}

First note that by $x \sqsubseteq_\alpha gx'$ and $x' \sqsubseteq_\alpha fx$,
also $x \sqsubseteq_\alpha hx$ and thus $y = h_\alpha x$ exists. 
Similarly, $y'= k_\alpha x'$ also exists.

Now by the 1st clause of Lemma~\ref{lem-falpha} 
$k y' \sqsubseteq_\alpha y'$, and since $g$ is $\alpha$-monotonic,
also $h gy' = gky' \sqsubseteq_\alpha gy'$. 
And since $x \sqsubseteq_\alpha gx'$ and $x' \sqsubseteq_\alpha y'$,
also $x \sqsubseteq_\alpha gy'$.  
We conclude by the 2nd clause of Lemma~\ref{lem-falpha} that $y \sqsubseteq_\alpha gy'$. 
Symmetrically, the same reasoning proves 
$y' \sqsubseteq_\alpha fy$.

Thus, $y \sqsubseteq_\alpha gy' \sqsubseteq_\alpha gf y = hy$.
But by Theorem~\ref{thm-fp}, it holds that $y =_\alpha hy$, so that  $y =_\alpha gy'$.
In a similar way, $y'=_\alpha fy$. 
This ends the proof of the lemma. 

We now return to the main proof. In order to 
show that $y_\alpha =_\alpha g y_\alpha'$ 
and $y'_\alpha =_\alpha fy_\alpha$ hold for $\alpha = 0$, note that 
$x_0 = \bot \sqsubseteq_0 g \bot_0'= g x_0'$, and symmetrically,
$x_0' = \bot' \sqsubseteq_0 f \bot = f x_0$. It follows by
Lemma~\ref{lem-zz} that $y_0 =_0 g y_0'$ and $y_0' =_0 f y_0$.

Suppose now that $\alpha > 0$ and our claim holds for all ordinals less than $\alpha$.
We distinguish two cases.

\emph{Case 1}: $\alpha = \gamma+ 1$ is a successor ordinal. 
Then, since the sequence $(y_\beta)_{\beta < \alpha}$ 
is compatible, by the induction hypothesis it holds 
that $x_\alpha = y_\gamma =_\gamma gy_\gamma' = gx_\alpha'$. 
But by the 3rd clause of Lemma~\ref{lem-falpha}, $y_\gamma$ is a 
$\sqsubseteq_\alpha$-least element of $[y_\gamma]_\gamma$,
hence $x_\alpha \sqsubseteq_\alpha gx_\alpha'$. Symmetrically,
$x'_\alpha \sqsubseteq_\alpha fx_\alpha$. 

\emph{Case 2}: 
$\alpha$ is a limit ordinal. Since the sequence $(y_\gamma)_{\gamma < \alpha}$ 
is compatible and hence increasing w.r.t. $\leq$, 
it holds that $x_\alpha = \bigvee_{\beta \leq \gamma < \alpha} y_\beta$
for all $\beta < \alpha$.
By compatibility, $y_\beta =_\beta y_\gamma$ for all $\beta \leq \gamma$, 
so that by Ax3, $x_\alpha =_\beta y_\beta$ for all $\beta < \alpha$.  
Symmetrically, $x'_\alpha =_\beta y'_\beta$, 
and since $g$ preserves the relation $=_\beta$, 
$gx'_\alpha =_\beta gy'_\beta$ for all $\beta < \alpha$. 
Also $y_\beta =_\beta gy'_\beta$  for all $\beta < \alpha$ by the induction hypothesis. 
This implies that $x_\alpha =_\beta y_\beta =_\beta gy'_\beta =_\beta gx_\alpha'$ for all $\beta < \alpha$.
Since $(y_\gamma)_{\gamma < \kappa}$ is a compatible sequence, 
by Lemma~\ref{lem-limitord}, $x_\alpha$ is the $\leq$-least and a $\sqsubseteq_\alpha$-least 
element of the set $(x_\alpha]_\alpha$. 
In particular, $x_\alpha \sqsubseteq_\alpha gx_\alpha'$. 
Symmetrically, $x'_\alpha \sqsubseteq_\alpha fx_\alpha$.

We have thus shown that in either case, $x_\alpha \sqsubseteq_\alpha gx_\alpha'$
and $x'_\alpha \sqsubseteq_\alpha fx_\alpha$. 
Thus, by Lemma~\ref{lem-zz}, $y_\alpha =_\alpha g y_\alpha'$ and 
$y_\alpha' =_\alpha f y_\alpha.$

Now by $y = \bigvee_{\alpha < \kappa} y_\alpha$ 
and $y' = \bigvee_{\alpha < \kappa} y'_\alpha$
and since $g$ is $\alpha$-monotonic, 
it holds that $y|_\alpha = y_\alpha$, $y'|_\alpha = y'_\alpha$, 
and $y =_\alpha y_\alpha =_\alpha gy'_\alpha =_\alpha gy'$ 
for all $\alpha < \kappa$. Thus, by $Ax2$, $h^\dagger = y = gy' = gk^\dagger$. 
Symmetrically, $k^\dagger = fh^\dagger$.

In order to establish the composition identity in its 
general form (\ref{eq-comp}), suppose now that $L,L',L''$ are models 
and $f: L \times L'' \to L'$ and $g: L' \times L'' \to L$
are $\alpha$-monotonic for all $\alpha < \kappa$. 
We want to show that (\ref{eq-comp}) holds. 
To this end, for every $z \in L''$, define $f_z: L \to L'$ and $g_z : L' \to L$ 
by $f_z x = f(x,z)$ and $g_z y = g(y,z)$ for all $x\in L$ and $y \in L'$. 
Then the functions $f_z$ and $g_z$ are also $\alpha$-monotonic for all $\alpha < \kappa$.
Moreover, since the parameter identity holds, 
\begin{eqnarray*}
(f\circ \langle g, \pi_{L''}^{L' \times L''}\rangle)^\dagger z 
&=& (f_z \circ g_z)^\dagger\\
(f \circ (\langle g\circ \langle f, \pi_{L''}^{L \times L''} \rangle)^\dagger)z
&=& f_z  (g_z \circ f_z)^\dagger. 
\end{eqnarray*} 
Since by the above argument  $(f_z \circ g_z)^\dagger =  f_z  (g_z \circ f_z)^\dagger$, hence 
\begin{eqnarray*}
(f\circ \langle g, \pi_{L''}^{L' \times L''}\rangle)^\dagger z 
&=& 
(f \circ (\langle g\circ \langle f, \pi_{L''}^{L \times L''} \rangle)^\dagger)z.
\end{eqnarray*} 
Since this holds for all $z$, we established the composition identity.

Next we prove that the double dagger identity holds. First let
 $f: L \times L\to L$ be $\alpha$-monotonic for all $\alpha < \kappa$,
where $L$ is a  model.  Since the fixed point identity holds, 
\begin{eqnarray*}
f \circ \Delta_L  \circ f^{\dagger\dagger} 
&=& 
f\circ \langle f^{\dagger \dagger},f^{\dagger\dagger} \rangle\\ 
&=& 
f \circ \langle f^\dagger \circ f^{\dagger \dagger}, f^{\dagger\dagger}\rangle \\
&=& 
f \circ \langle f^\dagger,\id_{L}\rangle \circ f^{\dagger\dagger} \\
&=& 
f^\dagger \circ  f^{\dagger\dagger}  \\
&=& f^{\dagger\dagger}.
\end{eqnarray*}
We conclude that $(f \circ \Delta_L  )^\dagger 
\sqsubseteq f^{\dagger\dagger}$.

Suppose now that $g: \mathbf{1} \to L$
and $$ f \circ \Delta_L  \circ  g 
= f \circ \langle g,g \rangle \sqsubseteq g.$$
We want to show that $f^{\dagger\dagger} \sqsubseteq g$. But
\begin{eqnarray*}
f \circ \langle g,g \rangle 
&=&
f \circ (\id_L \times g ) \circ g
\end{eqnarray*}
yielding
\begin{eqnarray*}
f \circ (\id_L \times g ) \circ g \sqsubseteq g.
\end{eqnarray*}
It follows that 
\begin{eqnarray*}
(f \circ (\id_L \times  g ))^\dagger \sqsubseteq g.
\end{eqnarray*} 
Thus, by the parameter identity 
\begin{eqnarray*}
f^\dagger \circ g \sqsubseteq g,
\end{eqnarray*} 
yielding $f^{\dagger\dagger} \sqsubseteq g$. Letting $g = (f\circ \Delta_L)^\dagger$,
we conclude that 
$f^{\dagger\dagger} \sqsubseteq (f\circ \Delta_L)^\dagger.$

Now for the general case, let $L$ and $L'$ be models and suppose that 
$f: L \times L \times L' \to L$ is $\alpha$-monotonic for all $\alpha < \kappa$.
Then $f^\dagger$ and $f \circ (\Delta_L \times \id_{L'})$ 
are $\alpha$-monotonic functions $ L \times L' \to L$ for all $\alpha < \kappa$.
Let $y \in L'$. We want to prove that 
$(f \circ (\Delta_L \times \id_{L'}))^\dagger y   =  f^{\dagger\dagger}y$.
But using the notation introduced above, 
$(f \circ (\Delta_L \times \id_{L'}))^\dagger y  = (f_y  \circ \Delta_L)^\dagger$ 
and $f^{\dagger\dagger}y = ((f^\dagger)_y)^\dagger = (f_y)^{\dagger\dagger}$,
moreover, $(f_y  \circ \Delta_L)^\dagger = (f_y)^{\dagger\dagger}$ by the previous case.

We still need to show that the weak functorial implication holds. 
Actually we will show that a stronger property holds. We will 
make use of the following concept. Suppose that $L$ and $L'$ are models, $\alpha < \kappa$ and $h: L \to L'$. 
We say that $h$ is strictly $\alpha$-continuous if it is $\alpha$-continuous, moreover, for each $x \in L$, 
$h(x|_\alpha) = (hx)|_\alpha$. Note that if $L' = L^n$, then the diagonal function $\Delta_L^n : L \to L^n$ 
is strictly $\alpha$-continuous for all $\alpha < \kappa$, since if $x \in L$ and $\alpha < \kappa$, 
then $\Delta_L^n(x|_\alpha) = (x|_\alpha,\ldots,x|_\alpha) = (x,\ldots,x)|_\alpha = (\Delta_L^n x)|_\alpha$.
Also note that $\Delta_L^n$ is continuous with respect to $\leq$ and preserves the 
least element.

{\em Claim. Suppose that $L,L'$ are models, $f: L \to L$ and $g: L' \to L'$
are $\alpha$-monotonic for all $\alpha < \kappa$ and $h: L \to L'$ is $\leq$-continuous 
and strictly $\alpha$-continuous for all $\alpha < \kappa$ and preserves the least element. 
Suppose that $h \circ f = g \circ h$. Then $f^\dagger = h \circ g^\dagger$.}

The proof of the claim relies on the explicit construction of $f^\dagger$ and $g^\dagger$. 
We will make use of the following lemma. 

\begin{lem}
\label{lem-uu}
Suppose that $x\in L$, $x' \in L'$ with $x \sqsubseteq_\alpha fx$ 
and $x' \sqsubseteq_\alpha gx'$, and let  $y = f_\alpha x$, $y' = g_\alpha x'$. 
If $x' = h x$ then $y' = h y$. 
\end{lem} 

In order to prove this lemma, we follow the construction of $f_\alpha x$ and $g_\alpha x'$. 
Let $x_0 = x$, and for each successor ordinal $\lambda = \delta + 1$, define 
$x_\lambda = f x_\delta$. When $\lambda$ is a limit ordinal, let 
$x_\lambda = \bigsqcup_\alpha \{x_\delta : \delta < \lambda\}$.
Define the sequence $(x'_\lambda)_\lambda$ in a similar fashion starting 
with $x'$ and using the function $g$. We prove by induction on $\lambda$ that 
$h x_\lambda = x'_\lambda$. 

When $\alpha = 0$, we have $h x_0 = h x = x' = x'_0$ by assumption.
Suppose now that $\lambda > 0$ and our claim holds for 
all ordinals less than $\lambda$. 

Let $\lambda$ be a successor ordinal, say $\lambda = \delta + 1$. 
Then $h x_\lambda = hf x_\delta = gh x_\delta = g x'_\delta = x'_\lambda$,
by the induction hypothesis. 
Suppose now that $\lambda$ is a limit ordinal. Then 
$h x_\lambda = h (\bigsqcup_\alpha \{ x_\delta : \delta < \lambda\})
=_\alpha \bigsqcup_\alpha \{ h x_\delta : \delta < \lambda\} 
= \bigsqcup_\alpha \{ x'_\delta : \delta < \lambda\}
= x'_\lambda$ by the induction hypothesis and since $h$ is $\alpha$-continuous.
But since $x_\lambda = x_\lambda|_\alpha$, $x'_\lambda = x'_\lambda|_\alpha$,
$hx_\lambda =_\alpha x'_\lambda$ and $h$ is strictly $\alpha$-continuous,
 it follows that $hx_\lambda = x'_\lambda$.

Since there is some ordinal $\lambda$ with $y = x_\lambda$ 
and $y'_\lambda = x'_\lambda$, the proof of the lemma is complete.

We now return to the proof of the claim. We know that $f^\dagger$ 
can be constructed as follows. We define $x_\alpha, y_\alpha \in L$ 
for $\alpha < \kappa$ by $x_\alpha = \bigvee_{\beta < \alpha} y_\beta$
and $y_\alpha = f_\alpha x_\alpha$. Define $x'_\alpha,y'_\alpha \in L'$
in a similar way using the function $g$.
Then $x_\alpha \sqsubseteq_\alpha fx_\alpha$ and 
$x'_\alpha \sqsubseteq_\alpha gx'_\alpha$ for all $\alpha < \kappa$,
moreover, $f^\dagger = y = \bigvee_{\alpha < \kappa} y_\alpha$ and $g^\dagger = 
y' = \bigvee_{\alpha < \kappa} y'_\alpha$. Since $y_\alpha \leq y_\beta$
and $y'_\alpha \leq  y'_\beta$ for all $\alpha < \beta < \kappa$, 
and since $h$ is $\leq$-continuous, it follows by $Ax2$ that 
$g^\dagger = h f^\dagger$ if we can show that 
$y' =_\alpha hy$ for all $\alpha < \kappa$. But for all $\alpha$, 
$y' =_\alpha hy$ holds iff $y'_\alpha =_\alpha h y_\alpha$, 
since $y =_\alpha y_\alpha$, $y' =_\alpha y_\alpha$ and $h$ 
is $\alpha$-continuous. 
Thus, $h f^\dagger = g^\dagger$ holds if $y'_\alpha =_\alpha h y_\alpha$
for all $\alpha < \kappa$. Actually we will prove that 
$y'_\alpha = h y_\alpha$ for all $\alpha < \kappa$. 

We prove by induction that $x'_\alpha = h x_\alpha$ and $y'_\alpha =
h y_\alpha$ for all $\alpha < \kappa$. 
We have $x'_0 = \bot' = h \bot = h x_0$, since $h$ preserves the 
least element, 
and thus $y'_0 = h y_0$ by Lemma~\ref{lem-uu}. Suppose now that 
$\alpha > 0$ and that our claim holds for all $\beta < \alpha$.
Now $x'_\alpha = \bigvee_{\beta < \alpha}  y'_\beta  
= \bigvee_{\beta < \alpha}   h y_\beta =
h(\bigvee_{\beta < \alpha} y_\beta) = h x_\alpha$
by the induction hypothesis and since $h$ is continuous.
Moreover, $y'_\alpha = h y_\alpha$, again by Lemma~\ref{lem-uu}. 

Suppose now that $L,L'$ are models and let $f: L \times L' \to L$ 
and $g: L^n \times L' \to L^n$ be $\alpha$-monotonic for all 
$\alpha < \kappa$ such that 
$\Delta_L^n\circ f =  g \circ (\Delta_L^n \times \id_{L'})$. 
Then for each fixed $y\in L'$, it holds that 
$\Delta_L^n\circ f_y =  g_y \circ \Delta_L^n$.
Thus, by the above claim, $\Delta_L^n(f_y)^\dagger = (g_y)^\dagger$,
i.e., $\Delta_L^n f^\dagger y = g^\dagger y$. 
Since this holds for all $y$, we conclude that 
$\Delta_L^n \circ f^\dagger = g^\dagger$. 
\eop

\begin{cor}
$\Mod_c$ is an iteration category with a weak functorial dagger. 
\end{cor}

By Remark~\ref{rem-complete lattices4}, 
an identity involving the cartesian operations and dagger holds 
in the category of complete
lattices and monotonic (or continuous) functions iff it holds 
in all iteration categories.  
Using this fact, Theorem~\ref{thm-iteration cat} 
and Example~\ref{expl-complete lattices3},
we obtain the following completeness
result.

\begin{cor}
\label{cor-exact} {\rm (Completeness)} 
The following conditions are equivalent for an identity 
$t = t'$ between terms involving the cartesian operations
and dagger:
\begin{itemize}
\item $t = t'$ holds in $\Mod_m$.
\item $t = t'$ holds in $\Mod_c$.
\item $t = t'$ holds in iteration categories. 
\end{itemize} 
\end{cor} 

{\sl Proof.} The fact that (i) implies (ii) is obvious.
By Example~\ref{expl-complete lattices3}, $\Mod_c$ 
contains the category of complete lattices 
and continuous functions equipped with the 
least fixed point operation as external dagger. Hence, 
by Remark~\ref{rem-complete lattices4}, any identity that
holds in $\Mod_c$ holds in iteration categories, 
proving that (ii) implies (iii). Finally, 
(iii) implies (i)  
by Theorem~\ref{thm-iteration cat}.
\eop

The same corollary may be derived from Theorem~\ref{thm-iteration cat} 
and a result proved in \cite{SimpsonPlotkin} showing that every
nontrivial iteration category having at least two morphisms 
$\one \to L$ for some object $L$ satisfies exactly 
the identities of iteration theories.

\begin{remark}
In \cite{Espark}, it is shown that least fixed points of \emph{monotonic} 
functions over partial ordered sets give rise to iteration theories even if 
the least fixed points cannot be constructed. 
This result is not applicable here, since our functions 
are not necessarily monotonic w.r.t. $\sqsubseteq$. Moreover, we have also established 
weak functoriality that does not necessarily hold for the least fixed point operation. 
\end{remark}

\section{Cartesian closed categories and the abstraction identity}
\label{sec-abstraction}

Following \cite{ERfp}, in this section we define certain stronger versions
of models that give rise to \emph{cartesian closed categories} (ccc's) \cite{BarrWells}. 
We establish the abstraction identity \cite{BEccc} that connects the fixed point operation 
to lambda abstraction. 

For any objects $L,L'$ in a ccc, we shall denote by $e_{L',L}$ 
an \emph{evaluation morphism} $(L' \to L) \times L'\to L$. 
Thus, for any $L,L',L''$ and $f: L' \times L'' \to L$,
there is a unique morphism $g : L'' \to (L' \to L)$ such that 
\begin{eqnarray*} 
f &=& e_{L',L} \circ (g \times \id_{L'}). 
\end{eqnarray*} 
Below we will denote this unique morphism $g$ by $\Lambda f$. 

We will consider ccc's equipped with an external dagger operation. 

\begin{remark} 
An external dagger operation over a ccc satisfying the parameter 
identity may be internalized and is determined by a family of 
morphisms $(L \to L) \to L$, where $L$ ranges over the objects.
For the internal forms of the identities mentioned in this
paper, we refer to \cite{BEccc}.  
\end{remark} 

We recall a new axiom from \cite{ERfp}. 

\emph{Ax5. For all $x_i,y_i$ with $x_i \sqsubseteq_\alpha y_i$ for all $i \in I$,
where $\alpha < \kappa$ and $I$ is an index set, it holds that $\bigvee \{x_i : i \in I\} 
\sqsubseteq_\alpha \bigvee \{y_i : i\in I\}$.}

Note that in all models, $Ax5$ implies $Ax4$. Moreover, 
$Ax5$ trivially holds in all models when $I$ is empty or a singleton set. 
Let $\Mod_m'$ denote the full 
subcategory of $\Mod_m$ spanned by those models satisfying 
$Ax5$. In \cite{CERhigher}, it is established in essence that 
$\Mod'_m$ is a cartesian closed category.

\begin{thm}
\label{thm-ccc1}
The category $\Mod_m'$ is cartesian closed. 
\end{thm} 

{\sl Proof.} 
It is clear that the product of any family 
of models satisfying $Ax5$ also satisfies this axiom. 
Since $\Mod_m$ is a cc, 
so is $\Mod_m'$. 

As expected, for any models $L,L'$ satisfying $Ax5$, 
the exponential object $(L' \to L)$ 
in $\Mod_m'$ is the collection of all functions 
$f: L' \to L$ which are $\alpha$-monotonic for all 
$\alpha < \kappa$.
Since each function $L' \to L$ may be seen as an element 
of the product $\prod_{x \in L'} L$, equipped with the (pre)orderings 
$\leq$ and $\sqsubseteq_\alpha$, $\alpha < \kappa$,  defined pointwise, 
$(L' \to L)$ is a model satisfying $Ax5$ provided that it is closed 
under the pointwise supremum operation $\bigvee$ w.r.t $\leq$
and the pointwise $\bigsqcup_\gamma$ operation, for all $\gamma < \kappa$. 

To prove this, let $G$ be a set of functions $L' \to L$ which are $\alpha$-monotonic 
for  all $\alpha < \kappa$. Define $f: L' \to L$ by $fx = \bigvee_{g \in G} gx$ for 
all $x \in L'$. Then $f$ is also $\alpha$-monotonic for all $\alpha < \kappa$.
Indeed, if $x \sqsubseteq_\alpha y$ in $L'$, where $\alpha < \kappa$, then 
$gx \sqsubseteq_\alpha gy$ for all $g \in G$ as each $g \in G$ 
is $\alpha$-monotonic.  Thus, since $Ax5$ holds in $L$, 
$fx = \bigvee_{g \in G} gx \sqsubseteq_\alpha \bigvee_{g \in G} gy = fy$. 

Suppose now that $\overline{g}: L' \to L$ is $\alpha$-monotonic for all 
$\alpha < \kappa$.  Let $\gamma < \kappa$, $G \subseteq (\overline{g}]_\gamma$
and $f = \bigsqcup_\gamma G$. We want to prove that $f$ is $\alpha$-monotonic
for all $\alpha < \kappa$. Suppose that $x \sqsubseteq_\alpha y$, where $\alpha < \kappa$. 
Since $G \subseteq (\overline{g}]_\gamma$, we have
 $\{gx : g \in G\} \subseteq (\overline{g}x]_\gamma$ 
and $\{gy : g \in G\} \subseteq (\overline{g}y]_\gamma$, 
so that $\bigsqcup_\gamma \{gx : g \in G\}$ and $\bigsqcup_\gamma \{gy : g \in G\}$
exist. Moreover, $fx = \bigsqcup_\gamma \{gx : g \in G\}$ and $fy = \bigsqcup_\gamma \{gy : g \in G\}$. 

If $\gamma <  \alpha$, then 
\begin{eqnarray*}
f x = \bigsqcup_\gamma\{gx : g \in G\} 
&=& \bigsqcup_\gamma\{gy : g \in G\} = fy
\end{eqnarray*}
since by $gx \sqsubseteq_\alpha gy$ and $\gamma < \alpha$ 
we have $gx =_\gamma gy$ for all $g\in G$.

Suppose now that $\gamma \geq  \alpha$. Then by Lemma~\ref{lem-assoc}
(or Lemma~\ref{lem-min}),   
\begin{eqnarray*}
(\bigsqcup_\gamma\{gx : g \in G\}\})|_\alpha 
  &=& \bigsqcup_\alpha \{gx : g \in G\}\\
&\sqsubseteq_\alpha & \bigsqcup_\alpha \{gy : g \in G\} \\
&=&  (\bigsqcup_\gamma\{gy : g \in G\})|_\alpha,
\end{eqnarray*}
since $gx \sqsubseteq_\alpha gy$ for all $g\in G$. 
Hence $(fx)|_\alpha \sqsubseteq_\alpha (fy)|_\alpha$, 
so that $fx \sqsubseteq_\alpha fy$. 
(Note that the assumption that $\overline{g}$ is $\alpha$-monotonic is used 
in both cases when $G$ is empty.) 
\eop

The evaluation map $e_{L',L}: (L' \to L) \times L' \to L$ is the usual 
evaluation function $(f,x) \mapsto fx$. For each $\alpha < \kappa$,
it is $\alpha$-monotonic in its first argument due to the pointwise
definition of the relation $\sqsubseteq_\alpha$ in $(L' \to L)$
and  $\alpha$-monotonic in its second argument since the 
functions in $(L'  \to L)$ are $\alpha$-monotonic. Finally,
when $f: L' \times L'' \to L$ is $\alpha$-monotonic for all 
$\alpha < \kappa$, then for all $y \in L''$, define $(\Lambda f)y : L' \to L$
by $(\Lambda f)yx = f(x,y)$ for all $x \in L'$. 
Then $(\Lambda f)y$ is $\alpha$-monotonic for all $\alpha < \kappa$ 
and 
$y \in L''$,  as is $\Lambda f$, since this holds for $f$. 
Indeed, if $\alpha < \kappa$, $y \in L''$ and $x \sqsubseteq_\alpha x'$ in $L'$, then 
$$
(\Lambda f)yx = f(x,y) \sqsubseteq_\alpha f(x',y) = (\Lambda f)yx'.
$$
And if $y \sqsubseteq_\alpha y'$ in $L''$ then 
$$(\Lambda f)y \sqsubseteq_\alpha (\Lambda f) y',$$
since for all $x$, $(\Lambda f)yx = f(x,y) \sqsubseteq_\alpha f(x,y')  = (\Lambda f) y'x$.
\eop

We will now define a cartesian closed subcategory of $\Mod_c$.
To this end, we introduce a new axiom. We say that $Ax6$ holds in a model
$L$ if $Ax5$ does and:

\emph{for all $\alpha < \kappa$, index set $I$ and nonempty linearly ordered 
set $(J,\leq)$, 
and for all $x_{i,j}\in L$ where $i \in I$ and $j \in J$ such that 
$x_{i,j} \sqsubseteq_\alpha x_{i,k}$ whenever $j \leq k$, it holds: 
\begin{eqnarray} 
\label{eq-ax6}
\bigvee_{i \in I} \bigsqcup_\alpha \{x_{i,j} : j \in J\}
&=_\alpha& 
\bigsqcup_\alpha \{ \bigvee_{i \in I} x_{i,j} : j \in J \}.
\end{eqnarray} }

\begin{remark}
Note that the above condition in $Ax6$ holds automatically when $I$ is empty, 
for in that case both sides of (\ref{eq-ax6}) are equal to $\bot$.
The axiom appears in \cite{ERfp}  
when $J$ is the linearly ordered set of nonnegative integers. 
\end{remark}

\begin{expl}
The standard model $V^Z$ satisfies Ax6. See also \cite{ERfp}. 
Also, each complete lattice $L$ viewed as a model as in Example~\ref{expl-complete lattices} 
satisfies $Ax6$. 
\end{expl} 

Let $\Mod_c'$ denote the full subcategory of $\Mod_c$ spanned by 
the models satisfying Ax6. 

\begin{thm}
The category $\Mod_c'$ is cartesian closed. 
\end{thm}

{\sl Proof.} 
First, if $L,L'$ are models, then the collection of functions $L' \to L$ 
which are $\alpha$-continuous for all $\alpha < \kappa$, equipped 
with the pointwise (pre)ordering relations $\leq$ and 
$\sqsubseteq_\alpha$, $\alpha  < \kappa$,
is also a model, giving rise to the exponential $(L' \to L)$.
This is in part due to the observation that 
when $G$ is a set of functions $g:  L' \to L$
which are $\alpha$-continuous for all $\alpha < \kappa$,
then the pointwise supremum $\bigvee G$ is also 
$\alpha$-continuous for all $\alpha < \kappa$. 
The proof of this fact uses Ax6. Indeed,
let $(I ,\leq)$ be a nonempty linearly ordered set 
and $x_i \in L'$ for all $i \in I$ such that $x_i \sqsubseteq_\alpha x_j$ 
whenever $i \leq j$ in $I$. Then, using $Ax5$ and $Ax6$ and the assumption that each $g\in G$ 
is $\alpha$-continuous, 
\begin{eqnarray*}
(\bigvee G)(\bigsqcup_\alpha \{x_i : i \in I\}) 
&=& 
\bigvee_{g \in G} g(\bigsqcup_\alpha \{x_i : i \in I\})\\
&=_\alpha& 
\bigvee_{g \in G} \bigsqcup_\alpha \{gx_i : i \in I\}\\
&=_\alpha&
\bigsqcup_\alpha \bigvee_{g \in G} \{ gx_i : i \in I \}\\
&=& 
\bigsqcup_\alpha \{ (\bigvee G)x_i : i \in I \}.
\end{eqnarray*}
(When $G$ is empty, both sides are equal to $\bot$.)

Suppose now that $\overline{g} : L' \to L$ is 
$\alpha$-continuous for all $\alpha < \kappa$
and $G \subseteq (\overline{g}]_\gamma$, where 
$\gamma < \kappa$ such that each $g \in G$ is 
$\alpha$-continuous for all $\alpha < \kappa$. 
We still need to prove that $f = \bigsqcup_\gamma G$ is 
$\alpha$-continuous for all $\alpha < \kappa$. 

Let $(I, \leq)$ be a nonempty linearly ordered
set and $x_i \in L'$ for all $i\in I$ 
such that $x_i \sqsubseteq_\alpha x_j$ whenever $i \leq j$  in $I$.
Then $gx_i \sqsubseteq_\alpha gx_j$ for all $g \in G$ and $i \leq j$ in $I$. 
Moreover, since $G \subseteq (\overline{g}]_\gamma$,
$\{gx_i : g \in G\}\subseteq (\overline{g}x_i]_\gamma$ for all $i\in I$. 
Hence, $\bigsqcup_\gamma \{gx_i : g \in G\} \in (\overline{g}x_i]_\gamma$
for all $i \in I$.

Let $f = \bigsqcup_\alpha G$. 
We want to prove that $f(\bigsqcup_\alpha \{x_i : i \in I\}) =_\alpha 
\bigsqcup_\alpha \{fx_i : i \in I\})$. 

First let $\gamma < \alpha$. 
Then
\begin{eqnarray*}
f(\bigsqcup_\alpha\{x_i : i \in I\})
&=& 
\bigsqcup_\gamma \{ g (\bigsqcup_\alpha \{ x_i: i \in I\}) : g \in G\}\\
&=& 
\bigsqcup_\gamma \{ \bigsqcup_\alpha \{ gx_i: i \in I\} : g \in G\}\\
&=& 
\bigsqcup_\gamma \{ gx_i : i \in I, g \in G\}
\end{eqnarray*}
where the third equality uses Lemma~\ref{lem-assoc} and the second equality 
is due to the fact that since $g(\bigsqcup_\alpha \{x_i : i \in I\}) =_\alpha 
\bigsqcup_\alpha \{gx_i : i \in I\}$ by $\alpha$-continuity, 
also $g(\bigsqcup_\alpha \{x_i : i \in I\}) =_\gamma
\bigsqcup_\alpha \{gx_i : i \in I\}$ by $\gamma < \alpha$ 
for all $g \in G$. On the other hand, 
\begin{eqnarray*}
\bigsqcup_\alpha \{fx_i : i \in I\}
&=& 
\bigsqcup_\alpha \{\bigsqcup_\gamma \{ gx_i : g \in G\} : i\in I\}\\
&=& 
\bigsqcup_\alpha \{\bigsqcup_\gamma \{ gx_i : g \in G, i\in I\}\}\\
&=& 
(\bigsqcup_\gamma \{ gx_i : g \in G, i\in I \})|_\alpha
\end{eqnarray*}
as $\bigsqcup_\gamma \{gx_i : g\in G\} = \bigsqcup_\gamma \{gx_j : g\in G\}$
for all $i = j$. (Hint: since $\gamma < \alpha$, $x_i =_\gamma x_j$ 
and $gx_i =_\gamma  gx_j$ for all $i,j\in I$ and $g \in G$.) 
But by Lemma~\ref{lem-min}, 
\begin{eqnarray*} 
(\bigsqcup_\gamma \{ gx_i : g \in G, i\in I\})|_\alpha
&=& 
\bigsqcup_\gamma \{ gx_i : g \in G, i\in I\},
\end{eqnarray*}
so that 
\begin{eqnarray*}
f(\bigsqcup_\alpha\{x_i : i \in I\})
&=& 
(\bigsqcup_\alpha \{fx_i : i \in I\})|_\alpha,
\end{eqnarray*}
ie., 
\begin{eqnarray*}
f(\bigsqcup_\alpha\{x_i : i \in I\})
&=_\alpha & 
\bigsqcup_\alpha \{fx_i : i \in I\},
\end{eqnarray*}

Next, suppose that $\gamma \geq \alpha$. Then using Lemma~\ref{lem-assoc} in the 
second, fourth and fifth lines and $\alpha$-continuity in the third, 
\begin{eqnarray*}
(f(\bigsqcup_\alpha\{x_i : i \in I\}))|_\alpha
&=& 
(\bigsqcup_\gamma \{ g (\bigsqcup_\alpha \{ x_i: i \in I\}) : g \in G\})|_\alpha\\
&= & 
\bigsqcup_\alpha \{g (\bigsqcup_\alpha \{ x_i: i \in I\}) : g \in G\}\\
&=& 
\bigsqcup_\alpha \{ \bigsqcup_\alpha \{gx_i : i \in I\} : g \in G\}\\
&=&
\bigsqcup_\alpha \{ \bigsqcup_\alpha \{ \bigsqcup_\gamma \{gx_i : i \in I\}\} : g \in G\}\\
&=& 
\bigsqcup_\alpha  \{ \bigsqcup_\gamma \{gx_i : i \in I\} : g \in G\}\\
&=& 
\bigsqcup_\alpha \{fx_i : i \in I\}, 
\end{eqnarray*}
so that $$(f(\bigsqcup_\alpha\{x_i : i \in I\}))|_\alpha = \bigsqcup_\alpha \{fx_i : i \in I\}.$$
Thus, $f(\bigsqcup_\alpha\{x_i : i \in I\}) =_\alpha
\bigsqcup_\alpha  \{fx_i : i \in I\}$ again.
(The assumption that $\overline{g}$ is $\alpha$-continuous has been used implicitly
when $G$ is empty.)

Now, for any pair of models $L,L'$ and any $\alpha < \kappa$, $\overline{f} : L' \to L$ 
in $\Mod_c'$ 
and for all $F \subseteq (L' \to L) \subseteq (\overline{f}]_\alpha$,  
the evaluation function $e = e_{L',L}$ satisfies 
$e(\bigsqcup_\alpha F, x) = \bigsqcup_\alpha \{e(f,x) : f \in F\}$ 
since $\bigsqcup_\alpha F$ is formed pointwise.  
And if $(J,\leq)$ is a nonempty linearly ordered set 
and $x_j \in L'$ for all $j \in J$ such that 
$x_j \sqsubseteq_\alpha x_k$ whenever $j \leq k$ in $J$,
and if $f \in (L' \to L)$, then $e(f,\bigsqcup_\alpha \{x_j : j \in J\})
=_\alpha \bigsqcup_\alpha \{e(f,x_j) : j \in J\}$ since $f$ is $\alpha$-continuous. 
Finally, if $f: L' \times L'' \to L$ is $\alpha$-continuous for all 
$\alpha < \kappa$, where $L,L',L''$ are models, then 
$(\Lambda f)y  : L' \to L$ is also $\alpha$-continuous for all $y\in L''$ 
and $\alpha < \kappa$ 
as is $\Lambda f: L'' \to (L' \to L)$. \eop

The abstraction identity was introduced in \cite{BEccc} 
in order to connect the Conway structure to 
exponentials in those ccc's which are Conway categories.

{\em Abstraction identity}
\begin{eqnarray*}
\Lambda(f^\dagger) &=& (\Lambda g)^\dagger
\end{eqnarray*}
where $f: L \times L' \times L'' \to L$ and 
$$g = f \circ ((\langle e_{L',L}, \pi^{(L' \to L) \times L'}_{L'}\rangle 
\circ 
\langle 
\pi^{L' \times (L'\to L)}_{(L'\to L)},
\pi^{L' \times (L'\to L)}_{L'}
\rangle) \times \id_{L''})
: L' \times (L' \to L) \times L'' \to L,$$
so that 
$$\Lambda g : (L' \to L) \times L'' \to (L' \to L).$$

Below we will establish the abstraction identity in $\Mod_m'$ and $\Mod_c'$.
We will make use of some further results including the fixed point induction rule 
of Theorem~\ref{thm-fpinduction}.

\begin{lem}
\label{lem-x}
Suppose that $L$ is a model and $f: L \to L$ is 
$\alpha$-monotonic for all $\alpha < \kappa$. 
Let $\alpha < \kappa$ and $z \in L$ 
with $fz \sqsubseteq_\alpha z$. Then 
either there
is some $\beta \leq \alpha$ with $f^\dagger \sqsubset_\beta z$,
or $f^\dagger =_\alpha z$.
\end{lem}

{\sl Proof.} Define $x_\alpha = \bigvee_{\gamma < \alpha} y_\gamma$ and 
$y_\alpha = f_\alpha x_\alpha$ for all ordinals $\alpha < \kappa$.
We know that $f^\dagger|_\alpha = y_\alpha$ for all $\alpha$. 
Thus, for all $\alpha < \kappa$ and $z\in L$ with $f(z) \sqsubseteq_\alpha z$, 
we have $f^\dagger \sqsubset_\alpha z$ iff 
$y_\alpha \sqsubset_\alpha z$ 
and $f^\dagger =_\alpha z$ iff $y_\alpha =_\alpha z$.

We prove by induction on $\alpha< \kappa$ that either there is some $\beta \leq \alpha$ 
with $y_\beta \sqsubset_\beta z$, or $y_\alpha =_\alpha z$. 
When $\alpha = 0$ this is clear, since $x_0 = \bot \sqsubseteq_0 z$, hence 
by Lemma~\ref{lem-falpha}, $y_0 \sqsubseteq_0 z$.

Suppose that $\alpha > 0$ and our claim holds for all ordinals less than $\alpha$. 
If there is some $\gamma < \alpha$ with $y_\gamma \sqsubset_\gamma z$ we are done. 
So without loss of generality we may assume that $y_\gamma =_\gamma z$ for 
all $\gamma < \alpha$. There are two cases.

Suppose first that  $\alpha$ is a successor ordinal, say $\alpha = \gamma + 1$.
Then $z \in [y_\gamma]_\gamma$, 
hence by the third clause of Lemma~\ref{lem-falpha}, $y_\gamma \sqsubseteq_\alpha z$. 
But $x_\alpha = y_\gamma$, thus $x_\alpha \sqsubseteq_\alpha z$, and since $f(z) \sqsubseteq_\alpha z$, 
we conclude that $y_\alpha \sqsubseteq_\alpha z$ by the second clause of Lemma~\ref{lem-falpha}. 

Suppose now that $\alpha$ is a limit ordinal. Since $y_\gamma =_\gamma z$ for all 
$\gamma < \alpha$, $x_\alpha = \bigvee_{\gamma < \alpha} y_\gamma \sqsubseteq_\alpha z$
by Lemma~\ref{lem-limitord}. Since $f(z) \sqsubseteq_\alpha z$, it follows 
by Lemma~\ref{lem-falpha} again that $y_\alpha \sqsubseteq_\alpha z$. 
\eop

If $L$ and $L'$ are models, we let $L^{L'}$ denote the model of all functions $L' \to L$ 
which is isomorphic to the $L'$-fold direct product of $L$ with itself. By 
Theorem~\ref{thm-fp} and Lemma~\ref{lem-x} we have:

\begin{cor}
\label{thm-fpinduction}
Suppose that $L$ and $L'$ are models and $f:L \times L' \to L$ is $\alpha$-monotonic 
for all $\alpha < \kappa$. If $g: L \times L' \to L$ 
is $\alpha$-monotonic for all $\alpha < \kappa$ such that 
$f \circ \langle g ,\id_{L'}\rangle \sqsubseteq g$ in the model $L^{L'}$,  
then $f^\dagger \sqsubseteq g$. 
\end{cor}

{\sl Proof.} 
Suppose that $g: L'\to L$ is $\alpha$-monotonic for all $\alpha < \kappa$ 
with $f \circ \langle g, \id_{L'}\rangle \sqsubseteq g$. If equality holds, then 
$f(gy,y) = y$ for all $y \in L'$, hence, $f^\dagger y \sqsubseteq gy$ for all $y \in L'$
by Theorem~\ref{thm-fp}. 
Thus for each $y$, either $f^\dagger y = gy$, or there is some $\alpha_y < \kappa$
with $f^\dagger y \sqsubset_{\alpha_y} gy$. When $f^\dagger y = gy$, define $\alpha_y = \kappa$, 
and then let $\alpha$ be the least ordinal in the set $\{\alpha_y : y \in L'\}$. 
If $\alpha = \kappa$ then $f^\dagger = g$, otherwise $f^\dagger \sqsubset_\alpha g$. 
In either case, $f^\dagger \sqsubseteq g$. 

Suppose now that $f \circ \langle g, \id_{L'}\rangle \sqsubset g$. Then 
by Remark~\ref{rem-prod}, there is 
some $\alpha < \kappa$ such that for all $y \in L'$ it holds that 
$f(gy,y) \sqsubseteq_\alpha y$, and there is some $y_0 \in L'$ with 
$f(gy_0,y_0) \sqsubset_\alpha y_0$. Now by Lemma~\ref{lem-x} (applied to $f_y$),
for each $y$ either $f^\dagger y =_\alpha gy$, 
or there is an ordinal $\beta_y \leq \alpha$ with $f^\dagger y \sqsubset_{\beta_y} gy$. 
Let $\beta = \min \{ \alpha, \beta_y : y \in L'\}$. Then $f^\dagger \sqsubset_\beta g$,
hence $f^\dagger \sqsubset g$. \eop

\begin{thm}
\label{thm-abstraction}
The abstraction identity holds in the categories $\Mod_m'$ and $\Mod_c'$.
\end{thm}

{\sl Proof.}  We prove this result only for morphisms $f: L \times L' \to L$
(i.e., when $L''$ is a singleton). In that case the assertion becomes $f^\dagger = (\Lambda g)^\dagger$,
where $$g = f \circ \langle e_{L',L}, \pi^{(L' \to L)\times L'}_{L'}\rangle 
\circ \langle \pi^{L' \times (L' \to L)}_{(L' \to L)},
\pi^{L' \times (L' \to L)}_{L'} \rangle  : 
(L'\to L) \to (L' \to L),$$
so that $\Lambda g : (L' \to L) \to (L' \to L)$. 

Notice that for all $f\in (L' \to L)$  and $y \in L'$,
$(\Lambda g)hy = f(hy,y)$, hence
$$ (\Lambda g)h = f \circ \langle h, \id_{L'}\rangle: L' \to L.$$
Thus we have
$$(\Lambda g)f^\dagger = f\circ \langle f^\dagger, \id_{L'}\rangle = f^\dagger,$$
since the fixed point identity holds. Suppose that $h \in (L' \to L)$ with 
$(\Lambda g)h \sqsubseteq h$. Then $f \circ \langle h, \id_{L'}\rangle \sqsubseteq h$,
hence $f^\dagger \sqsubseteq h$ by Corollary~\ref{thm-fpinduction}. 

We have proved that $f^\dagger$ is the least fixed point of $(\Lambda g)$ with respect to 
the ordering $\sqsubseteq$. Since $(\Lambda g)^\dagger$ is also a least fixed point, 
we conclude that $(\Lambda g)^\dagger = f^\dagger$. \eop

\section{Some variants of the categories} 
\label{sec-variations}

Several subcategories of $\Mod_m$ and $\Mod_c$ were 
introduced in \cite{CERhigher,ERfp,ERwollic}
in connection with logic programming and boolean grammars. 
In this section we mention some of them and establish that they 
are also cc's and/or ccc's equipped with an external dagger operation
satisfying the identities of iteration categories and, if applicable,  
the abstraction identity. 

Suppose that $L$ is a model. We say that $L$ is a strong model if it 
satisfies the following two axioms:

\emph{Ax7. For all $x,y \in L$ and $\alpha < \kappa$, if $x \leq y$ and $x =_\beta y$ 
for all $\beta < \alpha$, then $x\sqsubseteq_\alpha y$.}

\emph{Ax8. For all $x,y \in L$, if $x \leq y$ then $x|_\alpha \leq y|_\alpha$ 
for all $\alpha < \kappa$.}

\begin{expl}
Again, the standard model $V^Z$ satisfies these axioms as does every 
complete lattice as a model, cf. \ref{expl-complete lattices}. 
\end{expl}

\begin{remark}
In any model $L$, it holds that $x = \bigvee_{\alpha < \kappa} x|_\alpha$, for all $x \in X$.
It follows that if $x,y \in L$ with $x|_\alpha \leq y|_\alpha$ for all $\alpha < \kappa$,
then $x \leq y$. If Ax8 holds, then the converse is also valid, so that
for all $x,y \in L$, $x \leq y$ iff $x|_\alpha \leq y|_\alpha$ for all $\alpha < \kappa$. 

As noted in \cite{ERfp}, if a model $L$ satisfies Ax7, then the relation $\leq$ 
is included in the relation $\sqsubseteq$. Thus, in such models $L$, 
the greatest elements w.r.t. $\leq$ and $\sqsubseteq$ coincide. 
\end{remark}

\begin{prop}
Suppose that $L_i$ is a model for all $i \in I$ and let $L = \prod_{i \in I} L_i$.
Then $L$  satisfies Ax7 iff 
each $L_i$ does. Similarly, $L$ satisfies Ax8 iff each $L_i$ does. Thus, $L$ is a strong model iff
each $L_i$ is a strong model.
\end{prop}

{\sl Proof.} Immediate from the pointwise definition of the order relations in $L$. \eop 

\begin{prop}
Suppose that $L$ and $L'$ are models. If $L$ satisfies $Ax5$, $Ax7$ and $Ax8$, then so does 
$(L' \to L)$ in $\Mod_m$. Thus, if $L$ and 
$L'$ are strong models satisfying $Ax5$, then so is $(L' \to L)$ in $\Mod_m$.
\end{prop}

{\sl Proof.} This follows from the previous proposition 
and the fact that any product of models satisfying $Ax5$ 
also satisfies this axiom, since 
$(L' \to L)$ can be embedded in $\prod_{x \in L'} L$. \eop 

Similarly, we have:

\begin{prop}
Suppose that $L$ and $L'$ are models. If $L$ satisfies Ax6, Ax7 and Ax8, then so does 
$(L' \to L)$ in $\Mod_c$. Thus, if $L$ and 
$L'$ are strong models satisfying Ax6, then so is $(L' \to L)$ in $\Mod_c$.
\end{prop}

Let $\SMod_m$ and $\SMod_c$ denote the full subcategories of 
$\Mod_m$ and $\Mod_c$ determined by the strong models. 
Similarly, let $\SMod_m'$ and $\SMod_c'$ denote the full subcategories of 
$\Mod_m'$ and $\Mod_c'$ determined by the strong models. 

\begin{cor}
$\SMod_m$ and $\SMod_c$, equipped with the external dagger 
operations inherited from $\Mod_m$ and $\Mod_c$, respectively, are
iteration categories.
$\SMod_m'$ and  $\SMod_c'$ are ccc's satisfying the identities
of iteration categories and the abstraction identity. 
\end{cor} 

$Ax1$ and $Ax2$ are self-dual but $Ax3$ and $Ax4$ are not. Their duals are 
the following axioms.

\emph{Ax3d. For every $\alpha < \kappa$, $x \in L$  and $X\subseteq (x]_\alpha$ 
there exists some $z\in (x]_\alpha$ with the following 
properties: 
\begin{itemize}
\item $ z \sqsubseteq_\alpha X$, 
\item for all $y\in (x]_\alpha$, if $y \sqsubseteq_\alpha X$ then $y \sqsubseteq_\alpha z$ 
      and $y \leq z$.
\end{itemize}}

\emph{Ax4d. For every $\alpha < \kappa$ and nonempty $X \subseteq L$ and $y \in L$, 
if $X =_\alpha y$, then $\bigwedge X =_\alpha y$.}

The element $z$ is unique in Ax3d and we denote it by $\bigsqcap_\alpha X$. 
In particular, let $x|^\alpha = \bigsqcap_\alpha \{x\}$ for all $x \in X$
and $\alpha < \kappa$. 

Regarding $Ax7$ and $Ax8$, $Ax7$ is self-dual but $Ax8$ is not. 
The dual of $Ax8$ is:

$Ax8d$. For all $x,y \in L$, if $x \leq y$ then $x|^\alpha \leq y|^\alpha$ 
for all $\alpha < \kappa$.

We say that a model $L$ is a symmetric model if it also satisfies $Ax3d$ and $Ax4d$. 
A symmetric strong model is a symmetric model which is a strong model satisfying $Ax8d$. 
Let $\SymMod_m$ and $\SymMod_c$ denote the full subcategories of $\Mod_m$  
and $\Mod_c$ spanned by the symmetric models. Similarly, let $\SymSMod_m$ 
and $\SymSMod_c$ denote the full subcategories of $\Mod_m$ 
and $\Mod_c$ determined by the symmetric strong models.

\begin{cor}
$\SymMod_m$, $\SymMod_c$, $\SymSMod_m$ and $\SymSMod_c$,
 equipped with the external dagger 
operation inherited from $\Mod_m$ and $\Mod_c$, are cc's 
satisfying all identities of iteration categories.
\end{cor} 

Since the model obtained from a complete lattice as in 
Example~\ref{expl-complete lattices} is a symmetric strong model, 
it follows that an identity involving the cartesian operations
and dagger holds in any of the categories $\SMod_m'$, $\SMod_c'$, 
$\SymMod_m$, $\SymMod_c$, $\SymSMod_m$ and $\SymSMod_c$
iff it holds in iteration categories. 

One may also define a cartesian closed categories of symmetric 
models and symmetric strong models as subcategories of
$\Mod'_m$. These are ccc's and iteration categories 
and satisfy the abstraction identity. We skip the details. 

Similarly to Corollary~\ref{cor-exact}, each of the above categories 
satisfies an identity involving the cartesian operations and dagger 
iff the identity holds in all iteration categories.

\section{Conclusion}

We introduced several cartesian and cartesian closed categories of stratified 
complete lattices and equipped them with the stratified least fixed 
point operation as external dagger. 
We proved that all identities of iteration categories hold in 
these categories. And since each of these categories `extends' the category 
of complete lattices and monotonic or continuous functions 
equipped with the least fixed point operation as dagger, we concluded 
that an identity involving the cartesian operations and dagger 
holds in any of these categories iff it holds in all iteration categories. 
When the category is cartesian closed, we also established the 
abstraction identity. One may thus perform symbolic computations 
over these categories using \emph{the} standard equational properties
of fixed point operations.
This has significance for modular logic programming, see eg. 
\cite{LifschitzTurner}.

\end{document}